\documentclass[12pt,]{article}
\usepackage{amsmath}
\usepackage{amssymb}
\usepackage{latexsym}
\usepackage{epsfig}
\usepackage{graphicx}
\usepackage{psfrag}

\newcommand{\ntext}{}

\def\mh{\hat{m}}

\newcommand{\radius}{R}

\newcommand{\R}{{\mathbb R}}

\newcommand{\C}{{\mathbb C}}
\renewcommand{\S}{{\mathbb S}}

\def\hat{\widehat}
\def\tilde{\widetilde}
\def \bfo {\begin {eqnarray*} }
\def \efo {\end {eqnarray*} }
\def \ba {\begin {eqnarray*} }
 
\def \ea {\end {eqnarray*} }

\def \beq {\begin {eqnarray}}

\def \eeq {\end {eqnarray}}

\def \det {\hbox{det}}

\def \e {\varepsilon}

\def \p {\partial}

\def \a {\alpha}

\def\Z{{\Bbb Z}}

\def\p{\partial}

\def\R{\mathbb R}

\title{Isotropic transformation optics: \\ approximate  acoustic and quantum cloaking }

\author{Allan Greenleaf\footnote{Department of Mathematics, University of Rochester, Rochester, NY 14627}
, Yaroslav Kurylev\footnote{Department of Mathematical Sciences, University College London, Gower Str, London,
WC1E 6BT, UK} \\ Matti Lassas\footnote{Institute of Mathematics, Helsinki University of Technology, FIN-02015, Finland}, 
Gunther Uhlmann\footnote{Department of Mathematics, University of Washington, Seattle, WA 98195} \footnote{Authors listed in alphabetical order. AG and GU
are supported by US NSF, ML by Academy of Finland and YK by UK EPSRC.}}

\date{}

\begin{document}

\maketitle

\begin{abstract}

Transformation optics constructions have allowed the design of  electromagnetic,  acoustic 
and quantum parameters that steer  waves around
a  region without penetrating it, so that  the region is hidden from external observations. The material parameters are anisotropic, and singular at the interface between the cloaked and uncloaked regions, making physical realization a challenge.
 We address this problem
by  showing how to construct {\sl isotropic and nonsingular}
parameters that give {\sl approximate} cloaking to any desired degree of accuracy  for  
electrostatic, acoustic and quantum waves. The techniques used here may be  applicable to a wider range of transformation
optics designs.

For the Helmholtz equation, cloaking is possible outside a discrete set of frequencies or energies, namely the Neumann eigenvalues of the cloaked region.
For the frequencies or energies corresponding to the Neumann eigenvalues of the cloaked region, the ideal cloak supports trapped states; near these energies, an approximate cloak supports {\sl almost trapped states}.
This is  in fact  a useful feature,
  and we conclude by giving several quantum mechanical applications.
 
\end{abstract}

\section{Introduction}\label{sec-intro}

Cloaking devices designs based on
transformation optics  require anisotropic and singular\footnote{By {\it singular}, we mean that at least one of the eigenvalues  goes to zero or infinity at some points.}  material
parameters, whether the conductivity (electrostatic) \cite{GLU2,GLU3},   index of refraction (Helmholtz)
\cite{Le,GKLU1},
permittivity and permeability (Maxwell) \cite{PSS,GKLU1}, mass tensor (acoustic) 
\cite{ChenChan,Cummer2,GKLU5,Nor},  or
effective mass (Schr\"odinger)\cite{Zhang}. The same is true for other transformation
optics designs, such as those motivated by general relativity \cite{LePhil}; field rotators \cite{ChenChanRot}; concentrators \cite{Luo}; electromagnetic wormholes \cite{GKLU2,GKLU4}; or
beam splitters \cite{Rahm}. Both the anisotropy and singularity  present
serious challenges in trying to physically realize such theoretical plans
using metamaterials. In this paper, we give a  general method, 
{\sl isotropic transformation optics}, for dealing with both of these problems; 
we describe it in some detail in the context of cloaking, but it should also be
applicable to a wider range of transformation optics-based designs. 

A well known
phenomenon in effective medium theory is  that homogenization of isotropic
material parameters  may lead, in the small-scale limit, to anisotropic ones
\cite{Milt}. Using ideas from \cite{Ngu,Allaire,Cherka} and elsewhere,  we show how to
exploit this to find
cloaking material parameters that are at once both  isotropic and nonsingular, at the
price of replacing perfect cloaking with  \linebreak{\it approximate} cloaking (of arbitrary accuracy). 
This  method, starting with  transformation optics-based
designs and constructing  approximations to them, first by {\it nonsingular}, but still anisotropic, material parameters, and
then by nonsingular {\it  isotropic}  parameters,  seems to be a very flexible tool for
creating physically realistic theoretical designs, easier to  implement than the ideal ones due to the relatively tame nature of the materials needed,   yet essentially capturing the desired effect on waves. 

 In ideal cloaking, for any wave propagation governed by the Helmholtz equation at frequency $\omega$, there is a dichotomy \cite[Thm. 1]{GKLU1} between generic values of $\omega$, for which the waves must vanish within the cloaked region $D$, and the discrete set of Neumann eigenvalues of $D$, for which there exist {\sl trapped states}: waves which are zero outside of $D$ and equal to a Neumann eigenfunction within $D$.
In the approximate cloaking resulting from isotropic transformation optics that we will describe, trapped states for the limiting  ideal cloak give rise to {\sl almost} trapped states for the approximate cloaks. The existence of these should be considered as a feature, not a bug; we discuss this further in \linebreak Sec. \ref{sec almost trapped} and give applications in \cite{GKLU7}.

We start 
by considering isotropic  transformation optics for acoustic  (and hence, at frequency zero, electrostatic) cloaking. First recall ideal cloaking for the Helmholtz equation.
For a Riemannian metric $g=(g_{ij})$ in $n$-dimensional space, 
the Helmholtz equation with source term is 
\beq\label{one}
\frac1{\sqrt{|g|}} \sum_{i,j=1}^n \frac{\p}{\p x_i} \left(\sqrt{|g|} \, 
g^{ij} \frac{\p u}{\p x_j}\right) + \omega^2 u = p,
\eeq
where $|g|=\det(g_{ij})$ and $(g^{ij})=g^{-1}=(g_{ij})^{-1}$.
In the acoustic equation, for which ideal 3D spherical cloaking was described
by Chen and Chan \cite{ChenChan} and Cummer, et al., \cite{Cummer2},
$\sqrt{|g|} g^{ij}$ represents the anisotropic density and
$\sqrt{|g|}$ the bulk modulus. 

In \cite{GKLU1},  we showed that the  singular cloaking metrics $g$ 
for electrostatics constructed in \cite{GLU2,GLU3}, 
giving the same boundary measurements of electrostatic potentials as the Euclidian metric $g_0=(\delta_{ij})$, also
cloak with respect to solutions of the Helmholtz equation at any
nonzero frequency $\omega$ and with any source $p$. An example in 3D, with respect to
spherical coordinates $(r,\theta,\phi)$,  is
\beq\label{four}
(g^{jk})=
\left(\begin{array}{ccc}
2(r-1)^2\sin \theta & 0 & 0\\
0 & 2 \sin \theta & 0 \\
0 & 0 &  2 (\sin \theta)^{-1}\\
\end{array}
\right)
\eeq
on $B_2-B_1=\{1<r\le2\}$, with the cloaked region being the ball $B_1=\{0\le r\le1\}$.\footnote{ $B_R$ denotes the central ball of radius $R$. Note that  $\frac{\p}{\p\theta}, \frac{\p}{\p\phi}$ are not normalized to have length 1; otherwise, (\ref{four}) agrees with \cite[(24-25)]{Cummer2} and \cite[(8)]{ChenChan}, cf. \cite{GKLU5}.} This $g$ is the image
of $g_0$ under the singular transformation $(r,\theta,\phi)=F(r',\theta',\phi')$ defined by
$r=1+\frac{r'}2,\,\theta=\theta',\, \phi=\phi',\, 0<r'\le 2$,  which blows up the point
$r'=0$ to the {\it cloaking surface} $\Sigma=\{r=1\}$. The same transformation  was  used by
Pendry, Schurig and Smith \cite{PSS} for Maxwell's equations and gives rise to the cloaking structure that is referred to in \cite{GKLU1} as the {\emph{single coating}}. It was shown in \cite[Thm.1]{GKLU1} that if the cloaked region is given
any nondegenerate metric, then finite energy waves $u$ that satisfy the Helmholtz equation (\ref{one}) on
$B_2$ in the sense of distributions (cf. Sec. \ref{sec: perfect ac} below) have the same set of Cauchy data at $r=2$, i.e., the same acoustic boundary measurements,  as do the solutions for 
the Helmholtz equation for $g_0$ with source term $p\circ F$.
The part of $p$ supported within the cloaked region is undetectable at $r=2$,  while the part of $p$ outside $\Sigma$  appears to be shifted by the transformation $F$; cf. \cite{Zolla}. Furthermore, on the inner side
$\Sigma^-$ of the cloaking surface, the normal derivative of
$u$ must vanish, so that within $B_1$ the acoustic waves propagate as if $\Sigma$
were lined with a sound--hard surface. 
Within $B_1$, $u$ can be any Neumann eigenfunction; if $-\omega^2$ is {\sl not} an eigenvalue, then the wave must vanish on $B_1$, while if it  {\sl is}, then $u$ can equal any associated eigenfunction there, leading to what we refer to as a {\sl trapped state} of the cloak.

In Sec. \ref{sec: acoustics} we introduce isotropic transformation optics
in the setting of acoustics,  starting by  approximating the ideal  singular  anisotropic
density and bulk modulus by {\it nonsingular} anisotropic parameters.
Then,  using a homogeneization argument \cite{Ngu,Allaire}, we  approximate these nonsingular anisotropic  parameters by nonsingular {\it isotropic} ones. This yields almost, or approximate,  invisibility in the sense that the boundary observations for the resulting acoustic parameters converge to the corresponding ones for a homogeneous, isotropic  
medium.

In Sec. \ref{sec: schrodinger} we consider
the quantum mechanical scattering problem for the
time-independent Schr\"odinger equation at energy $E$,
\beq\label{scattering}
& &(-\nabla^2 +V(x) )\psi(x) =E\psi(x),\quad x\in \R^d,\\
& &\psi(x)=\exp(iE^{1/2}x\cdotp \theta)+\psi_{sc}(x),\nonumber
\eeq
where $\theta\in \R^d$, $|\theta|=1$, and $\psi_{sc}(x)$ satisfies
the Sommerfeld radiation condition. 
By a gauge transformation, we can reduce an \emph{isotropic} acoustic equation
to a Schr\"odinger
equation.
In this paper we restrict ourselves to the case when the potential $V$ is
 compactly supported, so that
\ba
\psi_{sc}(x)=\frac{a_V(E, {x}/{|x|},\theta)}{|x|^{\frac{d-1}2}}\cdot
e^{i\lambda |x|}+{\cal O}\Big(\frac{1}{|x|^{\frac{d}{2}}}\Big),\quad
\hbox{as }|x|\to \infty.
\ea
The function $a_{V}(E,\theta',\theta)$ is the {\emph scattering amplitude}
at energy $E$
of the potential $V$. The
inverse scattering problem consists of determination of $V$ from the
scattering amplitude. As $V$ is compactly supported,
this inverse problem is equivalent to the problem of determination of $V$
from boundary measurements. Indeed, if $V$ is supported in a domain $\Omega$,
we define the Dirichlet-to-Neumann (DN) operator $\Lambda_V(E)$ at energy $E$ for the potential $V$ as follows.
For any smooth function $f$ on $\p \Omega$, we set
\ba
\Lambda_{V}(E)f= \p_\nu \psi|_{\p \Omega}
\ea
where $\psi$ is the solution of the Dirichlet boundary value problem
\ba
(-\nabla^2+V)\psi=E\psi,\,\,  \psi|_{\p \Omega}=f.
\ea
(Of course, $\Lambda_V(E)=\Lambda_{V-E}(0)$.)
Knowing $\Lambda_V(E)$ is equivalent to knowing  $a_{V}(E,\theta',\theta)$ for all $(\theta',\theta)\in\S^{d-1}\times\S^{d-1}$.
Roughly speaking, $\Lambda_V(E)$ can be considered as knowledge
of all external observations of $V$ at energy $E$ \cite{Be}.

In Sec.  \ref{sec:sing qm} we also consider  the magnetic Schr\"odinger equation  with
magnetic potential $A$ and electric potential $V$,
\ba
(-(\nabla+iA)^2+V-E)\psi =0,\quad \psi|_{\p \Omega}=f,
\ea
which defines the DN operator, 
\ba
\Lambda_{V,A}(E)(f)=  \p_\nu \psi|_{\p \Omega}+ i (A\cdot \nu) f.
\ea

There is an enormous literature on unique determination of a potential,
whether
from scattering data or from boundary measurements of solutions
of the associated Schr\"odinger equation. 
In \cite{SyU} it was shown that an $L^\infty$ potential  is determined by
the
associated DN operator, and \cite{LaNa} and \cite{Chan} extended this to rougher
potentials.
In dimension $d=2$, it has been shown recently that uniqueness holds if $V$
is merely  in $L^p, p>2$ \cite{Bu}.

On the other hand, for  $d=2$ and each  $E>0$,  there are
continuous families of rapidly decreasing (but noncompactly supported)  potentials
which are \emph{transparent}, i.e., for which the scattering amplitude
$a_V(E,\theta',\theta)$ vanishes at a fixed energy $E$,
$a_V(E,\theta',\theta)\equiv a_0(E,\theta',\theta)=0$ \cite{GrinNov}. More recently,
\cite{HHLe}
described central potentials transparent on the level of
the ray geometry.

Recently, Zhang, et al., \cite{Zhang}  have described an ideal quantum
mechanical cloak
at any fixed energy $E$ and proposed a physical implementation.
The construction  starts with a homogeneous, isotropic mass
tensor $\mh_0$ and  potential $V_0\equiv 0$, and subjects this pair to the
same
singular
transformation (``blowing up a point") as was
used in
\cite{GLU2,GLU3,PSS}.
The resulting cloaking mass-density tensor $\mh$ and potential $ V$ yield a Schr\"odinger equation that is  the Helmholtz equation (at
frequency
$\omega=\sqrt{E})$ for the corresponding singular Riemannian metric, thus
covered by the analysis of  cloaking for the Helmholtz equation  in
\cite[Sec. 3]{GKLU1}. The cloaking mass-density tensor $\mh$ and potential are both singular, and $\mh$
infinitely anisotropic, at $\Sigma$, combining to make such a cloak difficult
to implement, with the proposal in \cite{Zhang} involving ultracold atoms trapped in an optical lattice.

In this paper, we consider the problem in dimension $d=3$. For each energy
$E$, we construct a family
$\{V_{n}^E\}_{n=1}^\infty$ of bounded potentials, supported in the annular
region $B_3-
B_1$, 
which  
act as an \emph{approximate
invisibility cloak}:
for any potential $W$ on $B_1$, the scattering amplitudes
$a_{V_{n}^E+W}(E,\theta',\theta)\to 0$ as $n \to\infty$. Thus, when surrounded by the cloaking potentials $V_n^E$,  the potential $W$ is
undetectable by waves at energy $E$,  asymptotically in $n$.
Furthermore, $E$ either is or is not a Neumann  eigenvalue for $W$ on the cloaked region. In the latter case, with high probability  the approximate cloak  keeps particles of energy $E$ from entering the cloaked region; i.e., the cloak is effective at energy $E$.  In the former case, the cloaked region supports ``almost trapped" states, accepting and binding such particles and thereby functioning as a new type of ion trap. Furthermore, the trap is magnetically tunable: application of a homogeneous magnetic field allows one to switch between the two behaviors \nolinebreak\cite{GKLU7}. 

In Sec. \ref{sec:sing qm}
we consider several applications to quantum mechanics of this approach. 
In the first, we study the magnetic Schr\"odinger equation and construct a family of
potentials which, when combined with a fixed homogeneous magnetic field, 
make the matter waves behave as if the potentials were almost
zero and the magnetic potential were blowing up near a point, thus giving the illusion of a locally singular magnetic
field. In the second, we 
describe ``almost trapped" states which are largely concentrated in the cloaked region.
For the third application, we use
the same basic idea of isotropic transformation optics but we replace
the single coating construction used earlier
by the double coating construction 
of \cite{GKLU1}, corresponding to metamaterials  deployed on both sides of the cloaking surface,  to make matter waves  behave as if 
confined to a three dimensional sphere, $\S^3$. 

Full mathematical proofs will appear elsewhere \cite{GKLU6}. The authors are grateful to A. Cherkaev and V. Smyshlyaev for useful 
discussions on homogenization,  to S. Siltanen for help with the numerics,
and to the anonymous referees for constructive criticism and additional references.

\section{Cloaking for the acoustic equation}\label{sec: acoustics}

\subsection{Background}

Our analysis is closely related to the inverse problem for electrostatics,  or Calder\'on's conductivity problem.
Let $\Omega\subset \R^d$ be a  domain, at the boundary of which electrostatic measurements are to be made,  and
denote by $\sigma(x)$  the anisotropic  conductivity within.
In the absence of sources, an electrostatic potential $u$ satisfies  a divergence form equation,
\begin{equation}\label{johty}
\nabla\cdot \sigma\nabla u = 0
\end{equation}
on $\Omega$. To uniquely fix the solution $u$ it is enough to give its value, $f$, on the
boundary.  {In the idealized case, one measures,  for all voltage distributions
 $u|_{\p \Omega}=f$ on the boundary the corresponding
current fluxes, $\nu\cdotp \sigma \nabla u$, 
where $\nu$ is the exterior unit normal to $\p\Omega$.} Mathematically 
this amounts to the knowledge of the Dirichlet--Neumann (DN) map,  $\Lambda_\sigma$.
corresponding to $\sigma$, i.e., the map taking the Dirichlet boundary
values of the solution to (\ref{johty}) to the
corresponding Neumann boundary values,
\begin{equation}
\Lambda_\sigma: \ \ u|_{\p \Omega}\mapsto \nu\cdotp \sigma \nabla u|_{\p \Omega}.
\end{equation}
 If $F:\Omega\to \Omega,\quad F=(F^1,\dots,F^d)$, is a
diffeomorphism with $F|_{\p \Omega}=\hbox{Identity}$, 
then by making the change of variables $y=F(x)$
and setting $u=v\circ F^{-1}$, we obtain
\begin{equation}\label{johty2}
\nabla\cdot \tilde \sigma\nabla v = 0,
\end{equation}
where $\tilde \sigma=F_*\sigma$ is the push forward of $\sigma$ in $F$,
\beq\label{eqn-transf law}
(F_*\sigma)^{jk}(y)=\left.
\frac 1{\det [\frac {\p F^j}{\p x^k}(x)]}
\sum_{p,q=1}^d \frac {\p F^j}{\p x^p}(x)
\,\frac {\p F^k}{\p x^q}(x)  \sigma^{pq}(x)\right|_{x=F^{-1}(y)}.
\eeq
This can be used to show that
\ba
\Lambda_{F_*\sigma}=\Lambda_\sigma.
\ea

Thus, there is a
large
(infinite-dimensional) family of conductivities which all give rise to the
same electrostatic
measurements at the boundary. This observation is due to Luc Tartar (see
\cite{KV2} for an account.)
Calder\'on's inverse problem  for anisotropic
conductivities is then the question of
whether two conductivities with the same DN operator must be 
push-forwards of each other. There are a number of positive results in this direction, but it was shown in \cite{GLU2,GLU3} that, if one allows  singular maps, then in fact there counterexamples, i.e., conductivities that are undetectable to electrostatic measurements at the boundary. See \cite{KSVW} for $d=2$.

>From now on,  for simplicity we will restrict ourselves to the three dimensional case. For each $R>0$, let   $B_R=\{|x|\le R\}$ and $\Sigma_R=\{|x|=R\}$ be the central ball and sphere of radius $R$, resp., in $\R^3$,  and let ${\it O}=(0,0,0)$ denote  the origin.
To construct an invisibility cloak, for simplicity we use the specific
singular coordinate transformation 
$F:\R^3- \{{\it O}\}\to \R^3- B_1,$
given by
 \beq\label{transformation}
x=F(y):=\left\{\begin{array}{cl} y,&\hbox{for } |y|>2,\\
\left(1+\frac {|y|}2\right)\frac{y}{|y|},&\hbox{for }0<|y|\leq 2. \end{array}\right.  
\eeq
Letting $\sigma_0=1$ be the homogeneous isotropic conductivity 
on $\R^3$, $F$ then defines a conductivity $\sigma$ on 
$\R^3-  B_1$ by the formula
\beq\label{eqn-transf law2}
\sigma^{jk}(x):=(F_*\sigma_0)^{jk}(x),
\eeq
{cf. (\ref{eqn-transf law}).
More explicitly, the matrix $\sigma=[\sigma^{jk}]_{j,k=1}^3$ is
\ba
\sigma(x) &=&2|x|^{-2}(|x|-1)^2\Pi(x)+ 2(I-\Pi(x)),\quad 1<|x|<2,
\ea
where $\Pi(x):\R^3\to \R^3$ is the projection to the radial direction, defined by
\beq\label{proj. formula}
\Pi(x)\,v=\left(v\,\cdotp \frac{x}{|x|}\right)\frac{x}{|x|},
\eeq
i.e., $\Pi(x)$ is represented by the matrix $|x|^{-2}xx^t$, cf.
\cite{KSVW}. }

One sees that  $\sigma(x)$ is singular,  as one 
of its eigenvalues, namely the one corresponding to the radial direction,
tends to $0$ as $|x|\searrow 1$.
We can then extend  $\sigma$ to $B_1$ as an arbitrary
smooth, nondegenerate (bounded from above and below) conductivity there.
Let $\Omega=B_3$; the conductivity $\sigma$ is then a cloaking conductivity on $\Omega$,
as it is indistinguishable from $\sigma_0$, {\it vis-a-vis} electrostatic
boundary measurements of electrostatic potentials, treated rigorously as bounded, distributional solutions of the degenerate elliptic boundary value problem corresponding to $\sigma$ \cite{GLU2,GLU3}.

The same construction of $\sigma|_{\Omega-B_1}$ was
proposed in Pendry, Schurig and Smith \cite{PSS}  to cloak the region $B_1$
from observation by electromagnetic waves at a positive frequency;
see also Leonhardt \cite{Le} for a related approach for Helmholtz in $\R^2$.

\subsection{Cloaking for Helmholtz: ideal  acoustic cloaks }\label{sec: perfect ac}

{The cloaking conductivity $\sigma$ above corresponds to a Riemannian metric $g_{jk}$ 
that is related to $\sigma^{ij}$ by
\beq \label{determinant}
\sigma^{ij}(x)= |g(x)|^{1/2} g^{ij}(x), \quad |g| 
= \left( \hbox{det}[\sigma^{ij}] \right)^2
\eeq
where $[g^{jk}(x)]$ is the inverse matrix of $[g_{jk}(x)]$ and
  $|g(x)|= \hbox{det}[g_{jk}(x)]$.
The Helmholtz equation, with source term $p$, corresponding to this cloaking metric  
has the form  
\beq\label{case 1}
& &\sum_{j,k=1}^3 |g(x)|^{-1/2}\frac \p{\p x^j}(|g(x)|^{1/2}g^{jk}(x) 
\frac \p{\p x^k} u)+\omega^2 u= p\quad\hbox{on }\Omega,\\
& &u|_{\p \Omega}=f. \nonumber
\eeq
For now, $g$ is allowed to be an arbitrary nonsingular Riemannian metric, $g_{inn}$, on $B_1$.
Reinterpreting the conductivity tensor $\sigma$ as a mass tensor ( which has the
same transformation law (\ref{eqn-transf law}) )  and $|g|^\frac12$ as the bulk
modulus parameter, (\ref{case 1}) becomes an acoustic equation,
\beq\label{case 3 B}
& &\left(\nabla \cdotp\sigma \nabla +\omega^2|g|^{\frac12} 
\right)u= p(x) |g|^{\frac12} \quad\hbox{on }\Omega,\\
& &u|_{\p \Omega}=f. \nonumber
\eeq

This is the form of the acoustic wave equation considered in \cite{ChenChan,Cummer2}; see also \cite{Cummer} for $d=2$, and  \cite{Nor} for a somewhat different approach.
As $\sigma$ is singular at the cloaking surface $\Sigma:=\Sigma_1=\p
B_1$, one has to carefully define what one means by ``waves", that is by  
solutions to (\ref{case 1}) or (\ref{case 3 B}). Let us  recall the precise definition
of the solution to (\ref{case 1}) or (\ref{case 3 B}), discussed in detail in 
\cite{GKLU1}.} We say that $u$
 is a \emph{finite energy}  solution of the Helmholtz equation (\ref{case 1}) or
the acoustic equation (\ref{case 3 B}) if 
\begin{enumerate}

\item
$u$ is {\sl square integrable} with respect to the metric, i.e., is in
the weighted
$L^2$-space,
\ba
u\in L^2_g(\Omega)=\{u:\  \|u\|^2_g:=\int_{\Omega}\,dx\,
 |g|^{1/2} |u|^2<\infty\};
\ea

\item 
the {\it energy} of $u$ is finite, 
\ba 
\|\nabla u\|_g^2:=\int_{\Omega}\,dx\,  |g|^{1/2} g^{ij}\p_i
u\p_j u<\infty;
\ea

\item
$u$ satisfies the Dirichlet boundary condition $u|_{\p \Omega}=f$; and

\item
 the  equation (\ref{case 3 B}) is valid in the
weak {\sl distributional} sense, i.e.,  for all ${ \psi} \in C^\infty_0(\Omega)$
\beq\label{cond 5 B}
\int_{\Omega}\, dx\, [-(| g|^{1/2} g^{ij}\p_i u) \p_j { \psi}+\omega^2 
u { \psi} |g|^{1/2}]
=\int_{\Omega}\, dx \,  p(x) { \psi} (x)
|g|^{1/2} .
\eeq

\end{enumerate}
This last can be interpreted as saying that any smooth superposition of point measurements of $u$ satisfies the same integral identity as it would for a classical solution.
We also note that, since $g$ is singular,  the term $ | g|^{1/2} g^{ij}\p_i u $
must also  be defined in an appropriate weak sense.

It was shown in \cite[Thm. 1]{GKLU1}  that if $ u$ is the finite energy 
solution of  the acoustic equation (\ref{case 3 B}), then 
$u(x)$ defines two functions $v^+(y), \, y \in \Omega$,
and $v^-(y), \, y \in B_1$, by the formulae
\beq\label{u-formula 1}
u(x)=\left\{\begin{array}{cl} v^+(y), &\hbox{where $x=F(y)$, for } 1<|x|<3, \\
 v^-(y),&\hbox{where $x=y$, for } 0<|x|<1.\end{array}\right.  
\eeq
These functions satisfy the following boundary value problems: 
\beq 
\label{u-formula 2}
(\nabla^2+\omega^2)v^+(y)&=&\tilde p(y):=p(F(y))\quad \hbox{in }\Omega,
\\ \nonumber
v^+|_{\p \Omega}&=&f,
\eeq
and 
\beq\label{u-formula 3}
(\nabla^2_{g_{inn}}+\omega^2|g_{inn}|^{1/2})v^-(y)&=&|g_{inn}|^{1/2}  p(y)\quad \hbox{in }B_1,
\\ \nonumber
\p_\nu v^-|_{\p B_1}&=&0
\eeq
where $\p_\nu u= \p_r u$ denotes the normal derivative on $\p B_1$.

In the absence of sources within the cloaked region, (\ref{u-formula 3}) leads, as mentioned in the introduction,  to the phenomenon of \emph{trapped states}: If $-\omega^2$ is \emph{not}  a Neumann eigenvalue for $(B_1,g_{inn})$, then $v^-\equiv 0$ on $B_1$, the waves do not enter $B_1$, and cloaking as generally understood holds. On the other hand, if $-\omega^2$\emph{ is} an eigenvalue, then $v^-$ can be any function in the associated eigenspace; indeed, one can have $v^+\equiv 0$, in which case the total wave $u$ behaves as a  bound state for the cloak, concentrated in $B_1$; for simplicity, we refer to this as a trapped state of the ideal cloak.

\subsection{Nonsingular approximate acoustic cloak }

Next, consider nonsingular approximations to the ideal cloak, which are
more physically realizable by virtue of having bounded anisotropy ratio; see 
\cite{RYNQ,GKLU3,KSVW} for analyses of cloaking from the point of view of similar truncations. Studying  the behavior of  solutions  to
the corresponding boundary value problems near the cloaking surface, as 
these nonsingular approximately cloaking conductivities tend to the ideal
$\sigma$, we will see that the Neumann boundary
condition  appears in (\ref{u-formula 3}) on the
cloaked region $B_1$.
At the present time, for  mathematical  proofs
\cite{GKLU6} of some of the results below we assume
that   $\sigma$ be chosen  to be  the  homogeneous,
isotropic conductivity, $\sigma=\kappa\sigma_0$ inside
$B_1$,
i.e.,  $\sigma^{jk}(x)=\kappa \delta^{jk}$,   with 
$\kappa\geq 2$ a constant  such that $-\omega^2$ is not a Neumann eigenvalue on $B_1$. The first assumption
is not needed for physical arguments, but  the second is.

To start, let $1<R<2$, $\rho:=2(R-1)$ and introduce
 the coordinate transformation 
 $F_R:\R^3- B_\rho\to \R^3- B_R$, 
 \ba
x:=F_R(y)=\left\{\begin{array}{cl} y,&\hbox{for } |y|>2,\\
\left(1+\frac {|y|}2\right)\frac{y}{|y|},&\hbox{for }\rho<|y|\leq 2. 
\end{array}\right.  
\ea
We define the  corresponding approximate conductivity, $\sigma_R$ as
\beq \label{R-ideal}
\sigma^{jk}_R(x)=\left\{\begin{array}{cl }\sigma^{jk}(x)
&\hbox{for } |x|>R,\\
\kappa \delta^{jk},&\hbox{for }|x|\leq R. \end{array}\right.  
\eeq
{\ntext Note that then $\sigma^{jk}(x)=
\left(\left(F_{R}\right)_*\sigma_0\right)^{jk}(x)$ for $|x|>R$, where
$\sigma_0\equiv 1$ is the homogeneous, isotropic conductivity (or mass density)
tensor,} Observe that, for each $R>1$, the conductivity $\sigma_R$ is nonsingular,
i.e., is bounded from above and below with, however, the lower bound going to $0$
as $R \searrow 1$. Let us define 
\beq \label{R-equation}
g_R(x)=\det(\sigma_R(x))^2=
\left\{\begin{array}{cl} 
1,&\hbox{for }|x|\geq 2,\\  
64|x|^{-4}(|x|-1)^4&\hbox{for } R<|x|<2,\\
\kappa^6,&\hbox{for }|x|\leq R, \end{array}\right.  
\eeq
cf. (\ref{determinant}). Similar to (\ref{case 3 B}),
 consider the solutions of 
\ba
(\nabla \cdotp\sigma_R \nabla +\omega^2 g_R^{1/2})u_R
&=&g_R^{1/2} p\quad\hbox{in }\Omega:= B_3\\
u_R|_{\p \Omega}&=&f,
\ea
As $\sigma_R$ and $g_R$ are now non-singular everywhere on $D$, 
we have 
the standard transmission conditions
on  $\Sigma_R:=\{x:\ |x|=R\}$, 
\beq\label{trans a1}
& & u_R|_{\Sigma_\radius+}=u_R|_{\Sigma_\radius-},\\  \nonumber
& &  
e_r\cdotp \sigma_R \nabla u_R|_{\Sigma_\radius+}=
 e_r\cdotp \sigma_R \nabla u_R|_{\Sigma_\radius-},
\eeq
where $e_r$ is the radial unit vector and $\pm$ indicates when the 
trace on $\Sigma_R$ is computed as the limit $r\to R^\pm$.

Similar to (\ref{u-formula 1}), we have
 \ba
u_R(x)=\left\{\begin{array}{cl} v_R^+(F_R^{-1}(x)),&\hbox{for } R<|x|<3,\\
 v_R^-(x),&\hbox{for } |x|\leq R,\end{array}\right.  
\ea
with $v_R^{\pm}$ satisfying
\ba
(\nabla^2+\omega^2)v_R^+(y)&=&p(F_R(y)) \quad \hbox{in
}\rho<|y|<3,\\ v_R^+|_{\p\Omega}&=&f,
\ea
and 
\beq \label{extra-equation}
(\nabla^2+\kappa^2\omega^2)v_R^-(y)&=&\kappa^2p(y), \quad \hbox{in }|y|<R.
\eeq
Next, using  spherical coordinates $(r,\theta,\varphi)$, $r=|y|$,
 the transmission conditions (\ref{trans a1}) on the 
surface $\Sigma_R$ yield
\beq \label{trans a2}
& &v_R^+(\rho,\theta, \phi)=v_R^-(R,\theta, \phi),
\\ \nonumber
& &\rho^2\, \p_rv_R^+(\rho,\theta, \phi)=\kappa R^2 \, \p_rv_R^-(R,\theta, \phi).
\eeq

Below, we are most interested in the case $p=0$, but 
 also analyze the case  
\beq\label{sum deltas}
p(x)=\kappa^{-2} \sum_{|\alpha|\leq N}q_\alpha \p_x^\alpha \delta_0(x),
\eeq
where $ \delta_0$ is the Dirac delta function at origin and
$q_\alpha\in \C$, i.e., there is a (possibly quite strong) point source
the cloaked region.  The Helmholtz equation (\ref{extra-equation}) on the
entire space
$\R^3$, with the above point source   and the standard
radiation condition, would give rise to
 the wave
\ba
u^p_0(y)=\sum_{n=0}^{N}\sum_{m=-n}^n p_{nm}h^{(1)}_n(\kappa \omega r)   
Y^m_n(\theta,\varphi),
\quad p_{nm}=p_{nm}(\omega),
\ea
where  $Y^m_n$ are spherical harmonics and $h_n^{(1)}(z)$ and $j_n(z)$ 
are the spherical Bessel functions, see, e.g., \cite{CK}.

In $B_R$ the function $v_R^-(y)$ differs  
from $u^p_0$ by a solution to the homogeneous equation
(\ref{extra-equation}), and thus for $r<R$
\ba
v_R^-(r,\theta,\varphi)&=&
\sum_{n=0}^\infty\sum_{m=-n}^n (a_{nm}j_n(\kappa
\omega r)+p_{nm}h^{(1)}_n(\kappa \omega r))   Y^m_n(\theta,\varphi),
\ea
with yet undefined $a_{nm}=a_{nm}(\kappa, \omega; R)$.
Similarly, for  $\rho<r< 3$,  
\ba
v^+_R(r,\theta,\varphi)&=&
\sum_{n=0}^\infty\sum_{m=-n}^n (c_{nm}h^{(1)}_n(\omega r)
+b_{nm}j_n(\omega r))Y^m_n(\theta,\varphi), 
\ea
with as yet unspecified $b_{nm}=b_{nm}(\kappa,\omega; R)$ and
$c_{nm}=c_{nm}(\kappa, \omega; R)$.

Rewriting the boundary value $f$ on $\p \Omega$ as
\ba
f(\theta,\varphi)
=\sum_{n=0}^\infty\sum_{m=-n}^n f_{nm}Y^m_n(\theta,\varphi),
\ea
we obtain, together with transmission conditions (\ref{trans a2}), the 
following equations for $a_{nm}, \, b_{nm}$ and $c_{nm}$:
\beq \label{system1}
& &    f_{nm}=
b_{nm}j_n(3\omega)+c_{nm} h_n^{(1)}(3\omega),\\ 
 \label{system2}
 & &   a_{nm} j_n(\kappa \omega R)+p_{nm} h^{(1)}_n(\kappa \omega R)  
=b_{nm}j_n(\omega
\rho)+c_{nm} h_n^{(1)}\omega \rho),\\
\label{system3}
& & \kappa R^2(\kappa \omega a_{nm}( j_n)'(\kappa \omega R)+\kappa
\omega p_{nm}(h^{(1)}_n)'(\kappa \omega R))\\ & & \quad \quad
=\rho^2(b_{nm} \omega(j_n)'(k \rho)+\omega c_{nm} (h_n^{(1)})'(\omega
\rho)).\nonumber
\eeq
  
When $\omega$ is not a Dirichlet  eigenvalue of the equation (\ref{case 3
B}), we can find the $a_{nm}$ and $c_{nm}$ from
(\ref{system2})-(\ref{system3}) in terms of $p_{nm}$ and $b_{nm}$, and
use the  solutions obtained and the equation (\ref{system1}) 
to solve for $b_{nm}$ in terms of $f_{nm}$ and  $p_{nm}$.
This   yields 
\beq
\label{asymptoticsn}
& &b_{nm}= \frac 1{
j_n(3\omega)+s_{n}h_n^{(1)}(3\omega)}
(f_{nm}-\tilde s_n h_n^{(1)}(3\omega) p_{nm}), \nonumber
\\
& &c_{nm}=
s_{n}b_{nm}-\tilde s_{n}p_{nm}
,\\
\nonumber
& &a_{nm}=t_nb_{nm}-\tilde t_n p_{nm}
\eeq
where
\ba
& &s_{n}=\frac{\kappa^2 R^2 j_n(\omega \rho) (j_n)'(\kappa  \omega R)-\rho^2
(j_n)'(\omega
\rho)  j_n(\kappa  \omega R)}
{\rho^2 (h_n^{(1)})'(\omega \rho)j_n(\kappa \omega R) -\kappa^2 R^2 h_n^{(1)}(\omega
\rho) (j_n)' (\kappa  \omega R)},
\\
& &
t_n=  \frac{\rho^2  j_n(\omega \rho)(h_n^{(1)})'(\omega\rho)
- \rho^2  (j_n)'(\omega \rho) h_n^{(1)}(\omega \rho)}
{\rho^2 (h_n^{(1)})'(\omega \rho)j_n(\kappa  \omega R) -\kappa^2 R^2 h_n^{(1)}(\omega
\rho)  (j_n)'(\kappa  \omega R)},\\
& &\tilde s_{n}=\frac{\kappa^2  R^2 h_n^{(1)}(\kappa  \omega R) (j_n)'(\kappa  \omega
R) -\kappa^2  R^2 (h_n^{(1)})'(\kappa  \omega R) j_n(\kappa  \omega R)}
{\rho^2 (h_n^{(1)})'(\omega \rho)j_n(\omega R) -\kappa^2 R^2 h_n^{(1)}(\omega \rho)
(j_n)' (\kappa  \omega R)},
\\
& &
\tilde t_n=  \frac{\rho^2  h_n^{(1)}(\kappa  \omega R)(h_n^{(1)})'(\omega\rho)
- \kappa^2  R^2(h_n^{(1)})'(\kappa  \omega R) h_n^{(1)}(\omega \rho)}
{\rho^2 (h_n^{(1)})'(\omega \rho)j_n(\kappa  \omega R) -\kappa^2  R^2 h_n^{(1)}(\omega
\rho) (j_n)'(\kappa  \omega R)}.
\ea
Recalling that $a_{nm},\, b_{nm}$ and $c_{nm}$ depend on $R$, 
let us consider what happens as $R\searrow 1$, i.e., as $\rho=2(R-1)\searrow 0$.
We use the asymptotics
\beq \label{gradshtein}
& &
j_n(\omega\rho) ={\it O}(\rho^n),\,\,
j_n'(\omega\rho)={\it O}(\rho^{n-1}); \\
\nonumber & &
h^{(1)}_n (\omega\rho) ={\it O}(\rho^{-n-1}),\,\, (h^{(1)}_n)' (\omega\rho) ={\it O}(\rho^{-n-2}),
\quad \hbox{as} \,\, \rho \to 0,
\eeq
and  obtain   
\beq\label {A eq}
& &s_{n}\sim \frac{c_1\rho^2  \rho^{n-1}+ c_2\rho^n}
{c_3\rho^2  \rho^{-n-2}+c_4 \rho^{-n-1}}\sim c_5\rho^{2n+1},\\
& &
t_n\sim  \frac{c_1'\rho^2 \rho^n \rho^{-n-2}+c_2' \rho^2 \rho^{n-1} \rho^{-n-1}}
{c_3'\rho^2 \rho^{-n-2} +c_4\rho^{-n-1}}\sim c'_5\rho^{n+1},\\
& &\tilde s_{n}\sim \frac{c_1''+ c_2''}
{c_3\rho^2  \rho^{-n-2}+c_4 \rho^{-n-1}}\sim c''_5\rho^{n+1},\\
& & \label {D eq}
\tilde t_n\sim  \frac{c_1'''\rho^2 \rho^{-n-2}+c_2''' \rho^{-n-1}}
{c_3'\rho^2 \rho^{-n-2} +c_4\rho^{-n-1}}\sim c'''_5,
\eeq
{\ntext 
assuming the constant $c_4$ does not vanish. 
The constant $c_4$ is the product of a non-vanishing constant
and $ (j_n)'(\kappa \omega)$. Thus the asymptotics (\ref{A eq})-(\ref{D eq}) 
are valid if  $-(\kappa \omega)^2$ is
not a Neumann eigenvalue of the Laplacian in  the cloaked region $B_1$
and $-\omega^2$ is not a Dirichlet eigenvalue of the Laplacian in  the domain  $\Omega$.
In the rest of this section
 we assume that this is the case.}

{
Since the  system (\ref{system1})-(\ref{system3}) is linear, we consider separately  
two cases, when $f_{nm} \neq 0,\,
p_{nm}=0$, and when $f_{nm}= 0,\,
p_{nm} \neq 0$.

In the case $f_{nm} \neq 0,\,
p_{nm}=0$,
we have
\ba
& &b_{nm}   = {\it O}(1), \quad
c_{nm}= {\it O}(\rho^{2n+1}), \\ \nonumber
& &a_{nm} = {\it O}(\rho^{n+1}), \quad
\hbox{as} \,\, \rho \to 0.
\ea
The above equations, together with (\ref{gradshtein}),  
imply that the wave $v_R^-$ in the approximately 
cloaked region $r < R$ tends to $0$ as $\rho \to 0$, with the term associated to the
spherical harmonic $Y_n^m$ behaving like ${\it O}(\rho^{n+1})$. As for the wave $v_R^+$
in the region $\Omega - B_R$, both terms associated to the
spherical harmonic $Y_n^m$ and involving $j_n(\omega r)$ and $h^{(1)}_n(\omega r)$,
respectively,  are of the same order
${\it O}(1)$ near $r=\rho$. However, the terms involving $h^{(1)}_n(\omega r)$ decay,
as $r$ grows,  becoming  ${\it O}(\rho^{2n+1})$ for $r\ge r_0>1$.

In the the second case,  when $f_{nm}= 0,\,
p_{nm} \neq 0$,
we see that 
\ba
a_{nm}\sim - \frac{h'_n(\kappa \omega R)}{j_n(\kappa \omega R)} p_{nm}
={\it O}(1), \quad \hbox{as} \,\, \rho \to 0.
\ea
Also,
\beq\label{bnm14}
b_{nm}={\it O}(\rho^{n+1}), \quad 
c_{nm}={\it O}(\rho^{n+1}), \quad \hbox{as} \,\, \rho \to 0.
\eeq
These estimates show that $v_R^+$ is of the order ${\it O}(1)$ near $r=\rho$. 
However, it decays as $r$ grows becoming ${\it O}(\rho^{n+1})$ for 
$r\geq r_0>1$.

Summarizing, when we have a source only in the exterior (resp., interior) of the
cloaked region, the effect in the interior (resp., exterior) becomes very small as
$R\to 1$. More precisely,
the solutions $v_R^\pm$ with converge to $v^\pm$, i.e.,
\ba
\lim_{R\to 1}v_R^\pm(r,\theta,\varphi)= v^\pm (r,\theta,\varphi),
\ea
where $v^\pm$ were defined in (\ref{u-formula 1}),  (\ref{u-formula 2}), and  (\ref{u-formula 3}). Equations (\ref{system2}),(\ref{asymptoticsn}) and (\ref{bnm14}) show how the Neumann boundary condition naturally appears
on the inner side $\Sigma^-$ of the cloaking surface.

\subsection{Isotropic nonsingular approximate acoustic cloak}\label{sec-approx second}

In this section we approximate the anisotropic approximate cloak 
$\sigma_R$ by isotropic conductivities, which then will themselves be approximate
cloaks. Cloaking by layers of homogeneous, isotropic EM media has  been 
proposed in  \cite{HFJ,ChenChan2}; 
see also \cite{Far} for a related anisotropic 2D approach based on homogenization. For general references on homogenization, see \cite{A,Ben,dM,Jik}; for some previous work on its application in the context of photonic crystals, see \cite{Guen1,Guen2,Guen3}.

We will consider the isotropic conductivities of the  form
\ba
\gamma_\e(x)=\gamma(x,\frac{r}{\e})
\ea
where $r:=r(x)=|x|$ is the radial coordinate, 
$\gamma(x,r')=h(x,r')I\in \R^{3\times 3}$ and $h(x,r')$  a smooth,
scalar valued function to be chosen later that is periodic in $r'$ with
period 1,
i.e., $h(x,r'+1)=h(x,r')$, satisfying $0<C_1\leq h(x,r')\leq C_2$.

Let
$s=(r,\theta,\phi)$ and $t=(r',\theta',\phi')$ be spherical coordinates
corresponding to
 two different scales.
Next we homogenize the conductivity
in the $(r',\phi',\theta')$-coordinates.
With this goal, we denote by $e^1=(1,0,0)$, $e^2=(0,1,0)$, and
$e^3=(0,0,1)$
the vectors corresponding to unit vectors in $r'$, $\theta'$ and
$\phi'$ directions,
respectively. Moreover,
let    $U^i(s,t),\, i=1, 2, 3,$ be
the solutions of
\beq\label{cell equation}
\hbox{div}_t(\gamma(s,t)(\hbox{grad}_t \cdot U^i(s,t)+e^i)=0,\quad
t=(r',\theta',\phi')\in \R^3,
\eeq
that are  $1$-periodic functions in $r',\theta'$ and $\phi'$ variables
that satisfy, for all $s$,
\ba
\int_ {[0,1]^3}\, dt'\, U^i(s,t')=0,
\ea
where, $t'=(r',\theta',\phi')$ and $dt =dr' d\theta' d\phi'$.

Define the  two-scale corrector matrices \cite{Ngu,Allaire,Allaire2} as
\ba
P_{j}^k(s,t)=\frac {\p}{\p t^j}U^k(s,t)+\delta^k_j.
\ea
Then the homogenized conductivity  is
\beq \label{corrector}
\hat \gamma^{jk}(s)=\sum_{p=1}^3
\int_ {[0,1]^3} \,dt\, \gamma^{jp}(s,t)P^k_p(s,t)
=\int_{[0, 1]^3}\, dt\, h(s, r') P^k_j(s,t),
\eeq
and satisfies $C_1I\leq \hat \gamma \leq C_2I$.

Since $\gamma$ is independent of $\theta',\, \phi'$,
the above condition implies that $U^i=0$ for $i=2,3$.
As for $U^1$, it satisfies
\bfo
\frac{\p}{\p r'} \left( h(s, r')\frac{\p U^1}{\p r'}\right)=
-\frac{\p  h(s, r')}{\p r'},
\efo
with $U^1$ being $1$-periodic with respect to $(\theta', \phi')$. These
imply that $U^1$ is independent of $(\theta', \phi')$ with
\bfo
\frac{\p U^1}{\p r'} =-1 + \frac{C}{h(s, r')}.
\efo
To find the constant $C$ we again use the periodicity of $U^1$,
now  with respect to $r'$, to get that $C$ is given by the harmonic means
$h_{harm}$ of $h$,
\beq \label{harmonic1}
C = h_{harm}(s):=\frac{1}{\int_0^1\, dr' \,  h^{-1}(s, r') }.
\eeq
Let  $h_a(s)$ denote the arithmetic means of $h$ in the second variable,
\ba
h_a(s) =\int_{[0,1]} \,dr'\, h(s,r').
\ea
Then the homogenized conductivity will be 
\ba
\hat \gamma(x)&=&h_{harm}(x)\Pi(x)+ h_a(x)(I-\Pi(x)),
\ea
where $\Pi(x)$ is the projection  
(\ref{proj. formula}).
For similar constructions see, e.g., \cite{Cherka}.

If 
\beq\label{Helmholtz}
& &(g_R(x)^{-1/2}\nabla \cdotp\gamma_\e(x) \nabla)w_\e= G\quad\hbox{on }\Omega,\\
& &w_\e|_{\p \Omega}=f, \nonumber
\eeq
applying results  analogous to  \cite{Ngu,Allaire} in spherical coordinates
 (see \cite{GKLU6}),
we  obtain
\beq \label{convergence 1}
\lim_{\e\to 0} w_\e = w, \quad \hbox{in} \,\, L^2(\Omega),
\eeq
where
\beq\label{Helmholtz B}
& &(g_R(x)^{-1/2}\nabla \cdotp \hat \gamma(x) \nabla)w= G\quad\hbox{on }\Omega,\\
& &w|_{\p \Omega}=f. \nonumber
\eeq
The convergence (\ref{convergence 1}) is physically reasonable; if we
combine spherical layers of conducting materials, the radial conductivity
is the harmonic average of the conductivity of layers and the tangential conductivity
is the arithmetic average of the conductivity of the layers. 
Applying this, the fact that $\hat \gamma$ and $\gamma_{\e}$ are 
uniformly bounded both from above and below, and results from the spectral theory, e.g., \cite{Kato}, one can show  \cite{GKLU6} that
if  
\beq\label{Helmholtz k2}
& &g_R(x)^{-1/2}\nabla \cdotp\gamma_{\e}(x) \nabla u_\e+\omega^2 u_\e = 
G\quad\hbox{on }\Omega,\\
& &u_\e|_{\p \Omega}=f \nonumber
\eeq
and $\omega^2$ is not a Dirichlet eigenvalue of the problem
\beq\label{Helmholtz B k2}
& &g_R(x)^{-1/2}\nabla \cdotp \hat \gamma(x) \nabla u+\omega^2u= G\quad\hbox{on
}\Omega,\\ & &u|_{\p \Omega}=f \nonumber
\eeq
then
\beq \label{convergence}
\lim_{\e\to 0} u_\e = u, \quad \hbox{in} \,\, L^2(\Omega).
\eeq

To consider an explicit isotropic conductivity, let us consider functions
\linebreak$\phi:\R\to \R$ and $\phi_L:\R\to \R$ given by 
\ba
\phi(t)=\left\{\begin{array}{cl}
0,& t<0,\\
\frac 12 t^2,& 0\leq t<1,\\
1- \frac 12 (2-t)^2,& 1\leq t<2\\
1,& t\geq 2,\end{array}
\right.
\ea
and
\ba
\phi_L(t)=\left\{\begin{array}{cl}
0,& t<0,\\
\phi(t),& 0\leq t<2,\\
1,& 2\leq t<L-2,\\
\phi(L-t),& t\geq L-2.\end{array}
\right.
\ea

 Let us use
\beq\label{gamma formula}
\gamma(r,\frac r\e)=\left[1+a^1(r)\zeta_1(\frac
r\e)-a^2(r)\zeta_2(\frac r\e)\right]^2, 
\eeq 
where
where  $a^2(r)$ is chosen positive and so that $\gamma(r, r') >0$
and,  for some positive integer $L$, we define $\zeta_j:\R\to \R$ to be $1-$periodic functions ,
\ba
\zeta_1(t)&=&\phi_L\Big(2Lt\Big),\quad 0\leq t<1,\\
\zeta_2(t)&=&\phi_L\Big(2L(t-\frac 12)\Big),\quad 0\leq t<1.
\ea

Temporarily fix an $R>1$; eventually, we will take a sequence of these $\searrow 1$.
In order to guarantee that the conductivity $\hat \gamma$ is smooth enough, we piece together
the cloaking conductivity in the exterior domain $r>R$ and
the homogeneous conductivity in the cloaked domain in a smooth manner.
To this end, we introduce a new parameter $\eta>0$
and solve for each $r$ the parameters $a^1(r)\geq 0$ and $a^2(r)\geq 0$ from the equations
for the harmonic and arithmetic averages,
\ba
& &\int_0^1 \,dr'\, [1+a^1(r)\zeta_1(r')-a^2(r)\zeta_2(r')]^{-2}\\
& &\ \ =
\left\{\begin{array}{cl}
2R^{-2}(R-1)^2(1-\phi(\frac{R-r}\eta))+ \kappa \phi(\frac{R-r}\eta),&\hbox{if } r<R,\\
2r^{-2}(r-1)^2,&\hbox{if }  R<r<2,\\
1,&\hbox{if } r>2,\end{array}\right.\\
& &
\int_0^1  \,dr'\,  [1+a^1(r)\zeta_1(r')-a^2(r)\zeta_2(r')]^2
\\
& &\ \ =
\left\{\begin{array}{cl}
2(1-\phi(\frac{R-r}\eta))+ \kappa \phi(\frac{R-r}\eta), &\hbox{if } r<R,\\
2, &\hbox{if } R<r<2,\\
1,&\hbox{if } r>2,\end{array}\right.
\ea
thus obtaining $a^1(r)$ and $a^2(r)$ such that the homogenized conductivity is
\beq\label{eqn gamma hat}
\widehat\gamma(x)=\sigma_{R,\eta}(x)=
\left\{\begin{array}{cl}
\pi_{R}(1-\phi(\frac {R-r}\eta))+ \kappa \phi(\frac {R-r}\eta),&\hbox{if } r<R,\\
\pi_R,&\hbox{if } R<r<2,\\
1,&\hbox{if } r>2,\end{array}\right.
\eeq
where
\ba
\pi_{R}=2R^{-2}(R-1)^2\Pi(x)+2(1-\Pi(x))
\ea
and  $\Pi(x)$ is as in
(\ref{proj. formula}). 
(In (\ref{eqn gamma hat}), the term $\kappa \phi(\frac{R-r}\eta)$
connects the exterior conductivity smoothly to the interior conductivity $\kappa$.)

We denote the solutions by
$a^1_{R,\eta}(r)$ and $a^2_{R,\eta}(r)$.
Now when first $\e\to 0$, then $\eta\to 0$ and finally $R\to 1$,
the obtained conductivities approximate better and better the cloaking conductivity $\sigma$.
Thus 
we choose appropriate sequences $R_n\to 1$, $\eta_n\to 0$ and $\e_n\to 0$ and denote
\beq\label{def gamman}
\gamma_n(x):=\left[1+a^1_{R_n,\eta_n}(r)\zeta_1(\frac r{\e_n})-a^2_{R_n,\eta_n}(r)
\zeta_2(\frac r{\e_n})\right]^2,\quad r=|x|. 
\eeq
Note also that if $a^1$ and $a^2$ are constant functions then $\gamma_n=
\gamma(x_0,x/\epsilon_n)$, so that all $\gamma_n$ look the ``same" inside the
 $\epsilon_n$ period; this is the case in Figs. 1 and 2. 
For later use, we need to assume that $\e_n$ goes  to zero faster than $\eta_n$, 
and so choose $\e_n<\eta_n^2$; we can also assume that all of the
$\e_n^{-1}\in\Z$, which ensures that the function $\gamma(x,r(x)/\e_n)$ is
$C^{1,1}$ smooth at $r=2$. Denoting
$g_n(x):=g_{R_n}(x)$, one can summarize the above analysis by:

\bigskip
\noindent
{\bf Isotropic approximate acoustic cloaking.}
{\it
If  $p$ is supported at the origin as in (\ref{sum deltas}),
then the solutions of

\beq\label{isotropic Helmoltz}
& &\left(g_n(x)^{-1/2}\nabla \cdotp\gamma_n(x) \nabla +\omega^2 \right)u_n=
p\quad\hbox{on }\Omega,\\ & &u_n|_{\p \Omega}=f, \nonumber
\eeq
tend to  a solution of (\ref{case 3 B}), as $n\to\infty$.}
\bigskip

This generalizes to the case when $p$ is a general source,
as long as its support does not intersect
the cloaking surface $\Sigma$, see  \cite{GKLU6}.

\section{Cloaking for the Schr\"odinger
equation}\label{sec: schrodinger}

\subsection{Gauge transformation}
This section is devoted to  approximate quantum cloaking at a fixed energy, i.e., for
the time-independent Schr\"odinger equation with the a potential $V(x)$,
$$
(-\nabla^2 +V)\psi =E\psi,\quad \hbox{in }\Omega.
$$

A standard  gauge transformation converts the equation (\ref{isotropic Helmoltz}) to such a Schr\"odinger equation. Assuming  that $u_n$ satisfies
equation (\ref{isotropic Helmoltz}) with $\omega^2=E$, and defining
\beq \label{gauge}
\psi_n(x)=\gamma_n^{1/2}(x) u_n(x),
\eeq
with $\gamma_n$ as in (\ref{def gamman}),
we then have that 
$$
\gamma_n^{-1/2} \nabla \cdotp\gamma_n \nabla(\gamma_n^{-1/2}\psi_n)=
\nabla^2\psi_n-\overline{V_n} \psi_n,
$$
where 
$$
\overline{V_n}=\gamma_n ^{-1/2}\nabla^2 \,
(\gamma_n^{1/2}).
$$
$\psi_n$ thus satisfies
 the  equation,
$$
(-\nabla^2+\overline{V_n} -E \gamma_n^{-1} g_n^{1/2})\psi_n=0 \quad \hbox{in
}\,\,\Omega,
$$
which can be interpreted as a Schr\"odinger at  energy $E$ by 
introducing  the effective potential
\beq
\label{effective-potential}
V_n^E(x):\,=\overline{V_n}(x)-E \gamma_n^{-1} g_{n}^{1/2}+E,
\eeq
so that
\beq \label{24.4.1}
(-\nabla^2+V_n^E)\psi_n=E\psi_n \quad \hbox{in }\,\, \Omega.
\eeq
We will show that  for generic $E$  the potentials $V_n^E$ function as approximate invisibility cloaks 
in quantum mechanics at energy $E$ (recall the discussion in the Introduction 
of the ideal quantum mechanical cloaking of \cite{Zhang}),
while for a discrete set of $E$, the approximate cloaks support {\sl almost} trapped states.

Let us next consider measurements made on $\p \Omega$.
Let $W(x)$ be a bounded potential supported on $B_1$, let 
$\Lambda_{W+V_n^E}(E)$ be the Dirichlet-to-Neumann (DN) operator
corresponding to the potential $W+V_n^E$, and $\Lambda_0(E)$ be
 the  DN operator, defined earlier, corresponding to the zero
potential.

The results for the acoustic equation given in Sec. 
\ref{sec: acoustics} yield the following result, constituting approximate 
cloaking in quantum
mechanics; for mathematical details of the proof, see \cite{GKLU6}.
\bigskip

\noindent
{\bf Approximate quantum cloaking.}
{\it Let $E \in \R$ be neither a Dirichlet eigenvalue of $-\nabla^2$ on $\Omega$
nor a Neumann eigenvalue of $-\nabla^2+W$ on $B_1$. 
 Then, the DN operators  (corresponding to boundary 
measurements at $\p\Omega$ of matter waves) for the  potentials
$W+V_n^E$ 
converge to the DN operator corresponding to free space, that is,
\ba
\lim_{n\to \infty} \Lambda_{W+V_n^E}(E) f= \Lambda_0(E) f 
\ea
in $L^2(\p \Omega)$, for any smooth $f$ on $\p\Omega$. 

Since convergence of the near field measurements imply convergence of 
the scattering amplitudes \cite{Be},
we also have
\ba
\lim_{n\to\infty} a_{W+V_n^E}(E,\theta',\theta)=a_0(E,\theta',\theta).
\ea}
\bigskip

Note that  the $V_n^E$
can be considered as almost transparent potentials at energy $E$, 
but this behavior is of a very different nature than the well-known results
from the classical theory of  spectral convergence, since the 
$V_n^E$ do {\it not} tend to $0$ as $n \to \infty$. (On the contrary, as we will see
shortly, they alternate and become unbounded near the cloaking surface $\Sigma$ as 
$n\to\infty$.) 
More importantly, the $V_n^E$ 
 also serve as  
approximate invisibility cloaks
for two-body scattering in quantum mechanics.  We note the following fundamental dichotomy:

{\bf Approximate cloaking vs. almost trapped states.} {\it  Any potential $W$ supported in $B_1$, when surrounded by $V_n^E$, becomes, for generic $E$, undetectable by matter waves at energy $E$, asymptotically in
$n$. Furthermore, the combination of  $\, W$
and the cloaking potential $V_n^E$ has negligible effect on waves passing the cloak.
On the other hand, for a discrete set of energies $E$,the potential  $W+V_n^E$ admits
\emph{almost trapped states}. This means that, if $E$ is an eigenvalue
of $W$ inside $B_1$, there are $E_n$ close to  $E$ 
which are  eigenvalues of $W+V_n^E$ in $\Omega$, and
the corresponding eigenfunctions are heavily concentrated in $B_1$;
see  Sec. 4.2 for details.}

As all measurement devices have limited precision, we can interpret this as saying that,
given a specific  device using particles  at energy $E$, one can design a potential to
 cloak an object from
any single-particle measurements made using that device.

\subsection{Explicit approximate quantum cloak}
We now make explicit the structure of the potentials $V_n^E$,  
obtaining analytic expressions used
to produce the numerics and figures below. 
Recall that the potential $\overline{V_n}$ when $\gamma_n$ is given by
(\ref{gamma formula}), with $L>4$ an integer. Since
\ba
\frac {d^2}{dt^2}\phi_L(t)=\left\{\begin{array}{rl}
0,& \hbox{if }t<0\hbox{ or }2\leq t<L-2\hbox{ or } L\leq t, \\
1,& \hbox{if }0\leq t<1\hbox{ or } L-1\leq t<L\\
-1,&\hbox{if } 1\leq t<2\hbox{ or } L-2\leq t<L-1\\
\end{array}
\right.
\ea
we see that
\beq\label{q formula}
\overline{V_n} &=&\gamma_n ^{-1/2}\nabla^2 \,
(\gamma_n^{1/2})\\ \nonumber
&=&
\e_n^{-2}\frac {a^1_{R_n,\eta_n}(r)\chi_n^1(\frac r {\e_n})-a^2_{R_n,\eta_n}(r)\chi_n^2(\frac r {\e_n})}
{1+a^1_{R_n,\eta_n}(r)\zeta_1(\frac r{\e_n})-a^2_{R_n,\eta_n}(r)\zeta_2(\frac
r{\e_n})}+{\cal O}(\e_n^{-1})
\eeq
where
\ba
\chi_n^1(r)=\left\{\begin{array}{rl}
1,& \hbox{if } r\in (0,1/L)+\Z\hbox{ and }R<r\e_n<2,\\
-1,& \hbox{if } r\in (1/L,2/L)+\Z\hbox{ and }R<r\e_n<2,\\
1,& \hbox{if } r\in ((L-2)/L,(L-1)/L)+\Z\hbox{ and }R<r\e_n<2,\\
-1,& \hbox{if } r\in ((L-1)/L,1)+\Z\hbox{ and }R<r\e_n<2,\\
0,& \hbox{otherwise},\end{array}
\right.
\ea
and $\chi_n^2(r)=\chi_n^1(r-\frac12)$.

We then see that the $\overline{V_n}$ are  centrally symmetric and can be considered
as being comprised of layers of   potential barrier walls and wells that become
very high and deep near the inner surface $\Sigma_{R_n}$. Each $\overline{V_n}$ is
bounded, but as
$n\to
\infty$, the height of the innermost walls and the depth of the innermost wells
goes to infinity when approaching the interface
$\Sigma$ from outside. These same properties are then passed from $\overline{V_n}$ to
$V_n^E$ by (\ref{effective-potential}). 

\vfil\eject
\begin{figure}[htbp]
\vspace{-4.8cm}
\begin{center}
\includegraphics[width=1.0\linewidth]{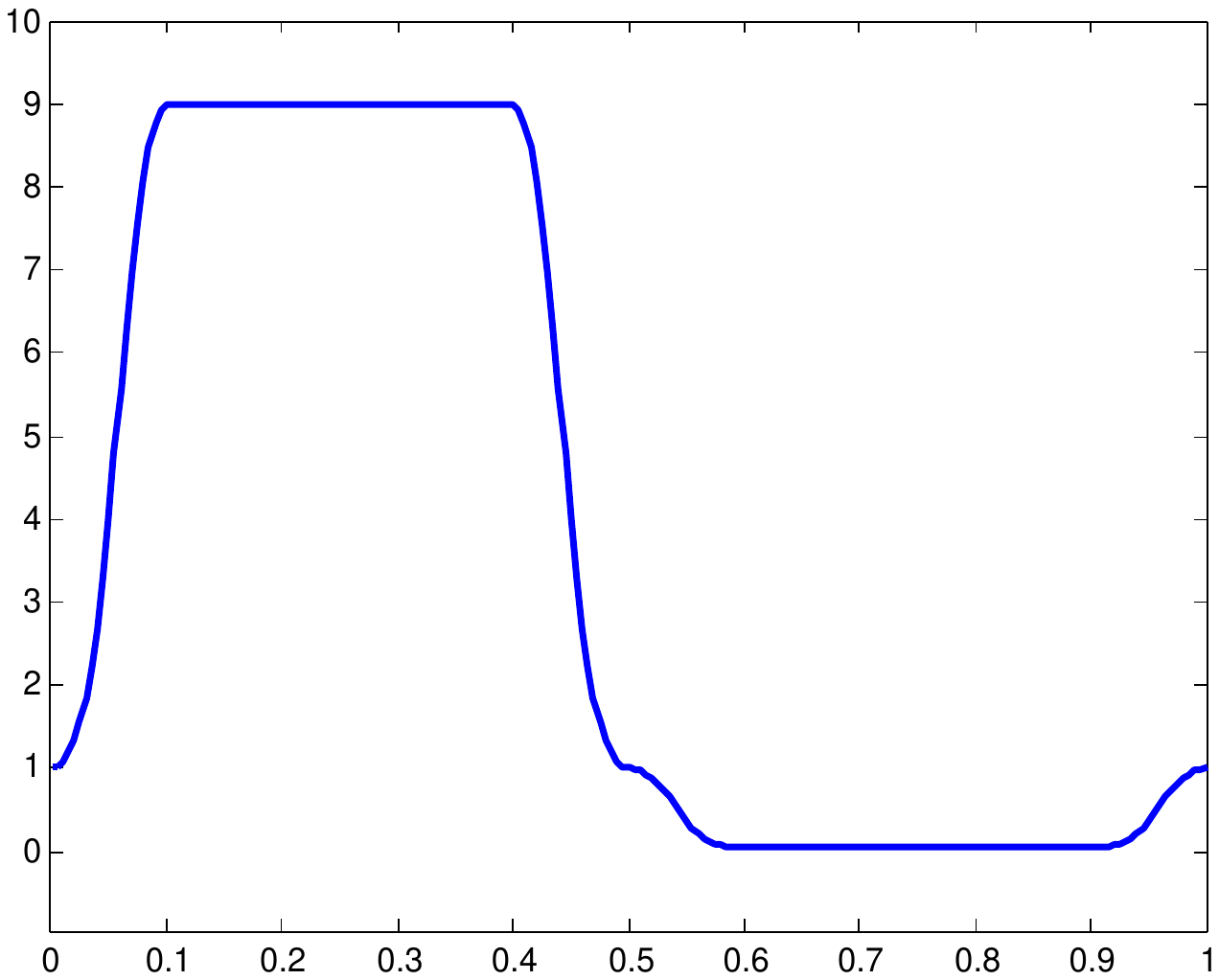}\qquad
\vspace{-5.0cm}
\end{center}
\caption{{ {\bf One radial cell of conductivity.} The isotropic conductivity $\gamma_n(x)$ is of the form
$\gamma_n(x)=h(x,\frac {|x|}{\e_n})$, where the function
$r'\mapsto h(x,r')$ with a fixed value of $x$ has period 1 in
variable $r'$. 
The horizontal axis  is $r'=\frac {|x|}{\e_n}\in [0,1]$.
The vertical axis is $h(x,r')$ as in (43), with the  $a^1=2$ and $a^2=0.8$. }}
\vspace{-1.0cm}
\end{figure}

\vfil\eject
\begin{figure}[htbp]
\vspace{-4.8cm}
\begin{center}
\includegraphics[width=1.0\linewidth]{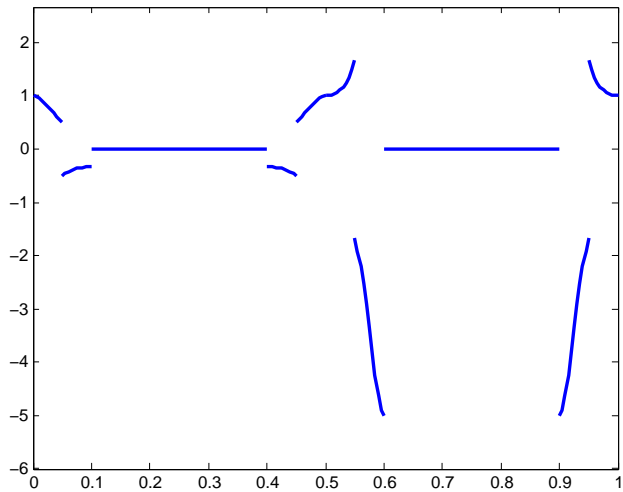}\qquad 
\vspace{-5.0cm}
\end{center}
\caption{ {\bf One radial cell of potential.} Potential $\overline{V}_n$ corresponding via (\ref{q formula}) to  conductivity $\gamma_n$ in  Fig 1.  The term of  order $\e_n^{-2}$, with the values $a^1=2$ and $a^2=0.8$,  is shown  as a function of  $r'=|x|/\e_n$,
as this  varies through the period $[0,1]$. }
\end{figure}

\vfil\eject
\subsection{Enforced boundary conditions on cloaking surface}

As  described in Sec. \ref{sec: perfect ac},, the natural boundary condition for the Helmholtz and
acoustic equations with perfect cloak, including those with sources within the cloaked region $B_1$, is the Neumann
boundary condition on $\Sigma^-$.\footnote{For  analysis of ideal cloaking, allowing various boundary conditions, as long as they are consistent with  von Neumann's theory of self-adjoint extensions, see Weder \cite{Weder}.}
However, the above analysis of approximate cloaking for
the  Schr\"odinger equation makes it possible to produce quantum cloaking devices
which enforce more general boundary conditions on $\Sigma^-$, e.g., the Robin
boundary conditions, which may be a useful feature in applications.

{To describe this, let
\ba
\chi_{{\tilde \e}}^0(|x|)=\left\{\begin{array}{cl}
1,& \hbox{if } 1-{\tilde \e}<r<1,\\
0,& \hbox{otherwise},\end{array}
\right.
\ea
with $\a =\a({\hat x}),\, {\hat x}= x/|x|$  a function on $\Sigma = \p B_1$.

Introduce an extra potential wall inside $B_1$ close to the surface
$\Sigma$, namely, take $W(x)$ in the form
\ba
W(x)=Q_{{\tilde \e}}(\a; |x|)=\a({\hat x})\frac{\chi_{{\tilde \e}}^0(|x|)}{{\tilde \e}},
\ea
and then consider the boundary value problem,
\beq\label{eqn vnm}
& &(-\nabla^2-E+V_n^E+Q_{{\tilde \e}})v=p\quad \hbox{in }B_3,\\
& &v|_{\p \Omega}=f.\nonumber
\eeq
As  $n\to \infty$ the solution $v=v_{n {\tilde \e}}$ to (\ref{eqn vnm}) tends,
inside $B_1$, to the
solution of
the equation
\beq\label{eqn vn}
& &(-\nabla^2-E+Q_{{\tilde \e}})v_{{\tilde \e}}=p\quad \hbox{in }B_1,\\
& &\p_r v_{{\tilde \e}}|_{\Sigma}=0.\nonumber
\eeq
Now, as ${\tilde \e} \searrow 0$,  we see that
\beq \label{delta-potential}
Q_{{\tilde \e}} \rightarrow \alpha\delta(r-1),
\eeq
so that
the functions $v_{{\tilde \e}}$ tend to the solution of the boundary value problem
\beq
\label{delta-eq}
\Big(-\nabla^2-E+\alpha\delta(r-1)\Big)v&=&p\quad \hbox{in }B_1,\\ \nonumber
\p_r v|_{\Sigma}&=&0.
\eeq
Note that to give the precise meaning of the above problem and its solution,
we should interpret (\ref{delta-eq})
in the weak sense.
Namely, $v$ is the solution to  (\ref{delta-eq}), if
for all ${ \varphi} \in C^\infty(B_1)$
\ba
\int_{B_1} \,dx\, [\nabla u\cdotp \nabla  \varphi-Eu\varphi ]\,+\,
\int_{\Sigma}\, dS(x)\, \alpha u \varphi\,
=\int_{B_1}\, dx\,  p { \varphi} 
,
\ea
which may be obtained from (\ref{delta-eq}) by a (formal) integration by parts
and utilizing (\ref{delta-potential}). However, the above weak formulation is
equivalent to the boundary value problem,
\ba
& &(-\nabla^2-E)v=p\quad \hbox{in }B_1,\\
& &\left(\p_r v-\alpha v\right)|_{\Sigma}=0.
\ea
Thus, the Neumann boundary condition
for the Schr\"odinger equation at the energy level $E$ has been replaced by a
Robin boundary condition on $\Sigma^-$, and the same holds for ideal acoustic cloaking.

Returning to  approximate cloaking, this means that if, for $\e,  {\tilde \e}$ very small, with $\e << {\tilde \e}$, we construct an approximate cloaking
potential with layers of thickness $\e$ and height $\e^{-1}$, and augment it by
an innermost  potential wall of width
${\tilde \e}$ and height ${\tilde \e}^{-1}$, then we obtain an approximate
quantum cloak with the wave inside $B_1$ behaving as if it satisfies the
Robin boundary condition.
It is clear from the above that the boundary condition appearing on the
cloaking surface is very dependent on the fine structure of the approximately cloaking  potential.
Physically, this boundary condition may be enforced by appropriate design of this extra
potential wall (rather than being due  to the cloaking
material in $B_3 - B_1$), so that we refer to this as an
{\it enforced boundary condition} in approximate cloaking, as opposed to the natural
Neumann condition that occurs in ideal cloaking.

\vfil\eject

\subsection{Approximation of $V^E_n$ with point charges}

One possible path to physical realization of these approximate quantum mechanical
cloaks  would be via electrostatic potentials,  approximating (again!) the potentials
$V^E_n$
by sums of point sources. Indeed, solving  the equation
\ba
V^E_n(x)=\int_{B_{R_\infty}}\, dy\,  \frac {-f^E_n(y)}{2\pi|x-y|}\,,\quad x\in \R^3, \ R_\infty>>1.
\ea
for  $f_n^E$ is an  ill-posed problem, but
using regularization methods one could find  approximate
solutions; the  resulting $f^E_n(x)$ could then be
approximated by a sum of delta functions, giving  blueprints   
for approximate cloaks implemented by electrode arrays.

\subsection{Numerical results}

We use
the analytic expressions found above to compute the fields
for a plane wave with $E_{in}(x)=Ae^{ikx\cdotp \vec d}$. 
The computations are made without reference to physical units;
 for simplicity, we use
$E=0.5$, $\kappa=2$ and  amplitude $A=1$.
Unless otherwisely stated, the cloak has parameters $\rho=0.01$,
i.e. $R=1.005$,   so that the anisotropy ratio \cite{GKLU3} is $4\times 10^4$,
and $\eta=0.055$. 
 In the simulations we use a cloak consisting of 20 homogenized
layers inside and 30 homogenized layers outside of the cloaking surface
$\Sigma_R=\{r=R\}$.
This means that $\epsilon$ inside the cloaking surface is $\eta/20$ and
outside the cloaking surface $(2-R)/30$.

Inside the cloak we have located a spherically symmetric potential;
$$W(x)=v_{in}\chi_{[0,0.9]}(r).$$ 
To illustrate
approximate cloaking, we used $v_{in}=-98$; to obtain an
almost trapped state,
$v_{in}=2.36$.

In the numerical solution to obtain the solutions
$\psi_n$ and $u_n$ we use the approximation that $L>>1$.
This implies 
that the cloaking conductivity $\gamma_R$ is piecewise constant, and
correspondingly,
 the cloaking potential $V_n^E$ is a weighted sum of delta functions,
and their derivatives, 
on spheres. In the numerical solution of the problem,
we represent the solution $u_n$ of the equation
$\nabla\cdotp \gamma_n\nabla u+g_n^{1/2}\omega^2u=0$
 in terms of Bessel functions up to order
$N=14$ in each
layer where the cloaking conductivity is constant. The 
transmission condition on the boundaries of these layers are
solved numerically by solving linear equations.
After this we compute the solution $\psi_n$ of the 
Schr\"odinger equation using formula $\psi_n(x)=\gamma_n(x)^{1/2}u_n(x)$.

Below we give the numerically computed coefficients of spherical harmonics
$Y^n_0$ in the case when $v_{in}=-98$ and $\rho=0.01$, in which we do not
have an eigenstate inside the cloaked region. The result are
compared to the case when we have scattering from the potential $W$
without a cloak. 

\vspace{2.0cm}

\centerline{\bf Table 1. coefficients of scattered waves for
 $v_{in}=-98$ and $\rho=0.01$}
\medskip
\centerline{\begin{tabular}{ccc}
\hline
$n$ &    $c_n$ with cloak and $W$ &   $c_n$ with $W$ but no cloak \\
\hline
0 &  $-0.0057 - 0.0751i $  &  $  + 0.8881i$     \\
1 &  $+0.0107 - 0.0000i $  &  $  - 0.0592i$     \\
2 &  $+0.0000 + 0.0052i $  &  $  - 0.1230i$     \\
3 &  $-0.0007 + 0.0000i $  &  $  - 0.0153i$     \\
4 &  $-0.0000 - 0.0000i $  &  $  + 0.0011i$     \\
5 &  $+0.0000 - 0.0000i $  &  $  + 0.0000i$     \\
6 &  $+0.0000 + 0.0000i $  &  $  + 0.0000i$     \\
\hline\\
$(\sum |c_n|^2)^{1/2}$  & $   0.0058  $ &   0.8076 \\
\end{tabular}}

\bigskip

\begin{figure}[htbp]
\begin{center}
\vspace{-6.0cm}
\centerline{\includegraphics[width=1.2\linewidth]{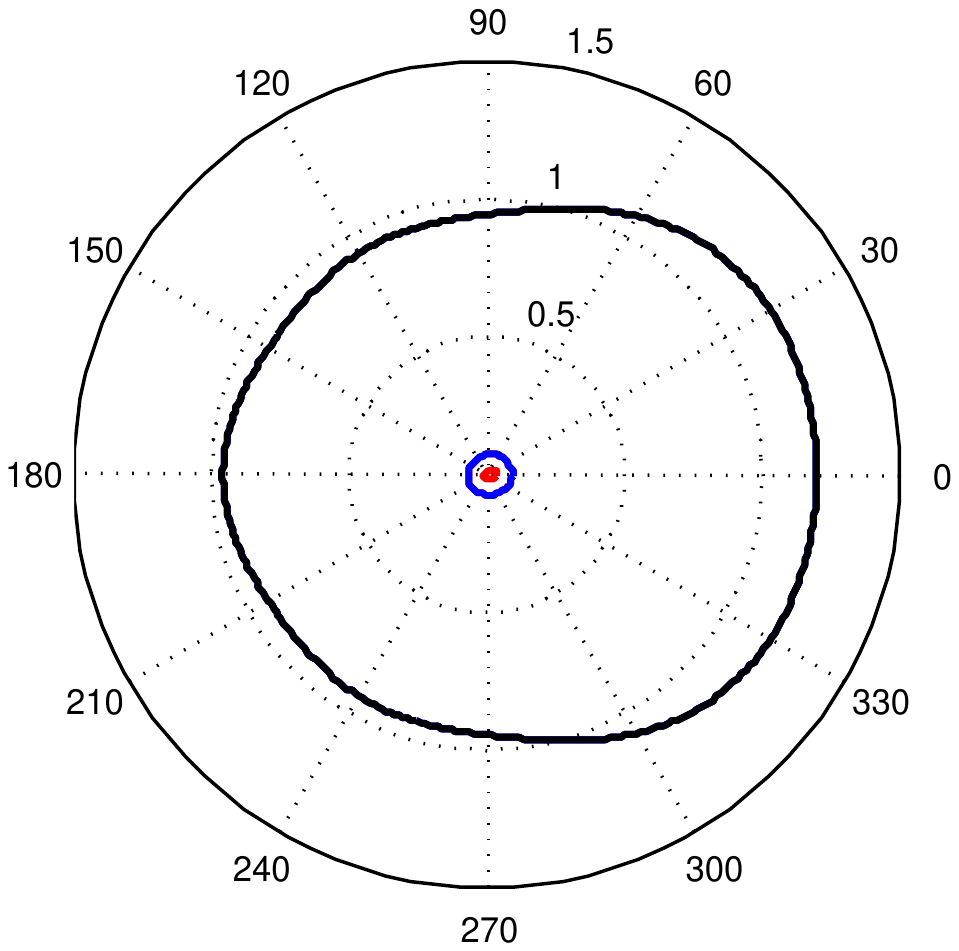}}
\end{center}
\vspace{-7.0cm}
\caption{ {\bf The magnitudes of the far fields.} 
The far-fields $\theta\mapsto |a(\theta,\varphi)|$ with 
$\varphi=0$ are shown: {\bf Black curve:} scattering
from $W$ without the  cloak.  {\bf Blue curve:} scattering from $W$ surrounded by
cloak,  $\rho=0.1$; \qquad\qquad\qquad
{\bf Red curve:}  scattering from $W$ 
surrounded by cloak, $\rho=0.01$.}
\end{figure}

\vfil\eject

\begin{figure}[htbp]
\begin{center}
\vspace{-7cm}
\centerline{
\includegraphics[width=0.83\linewidth]{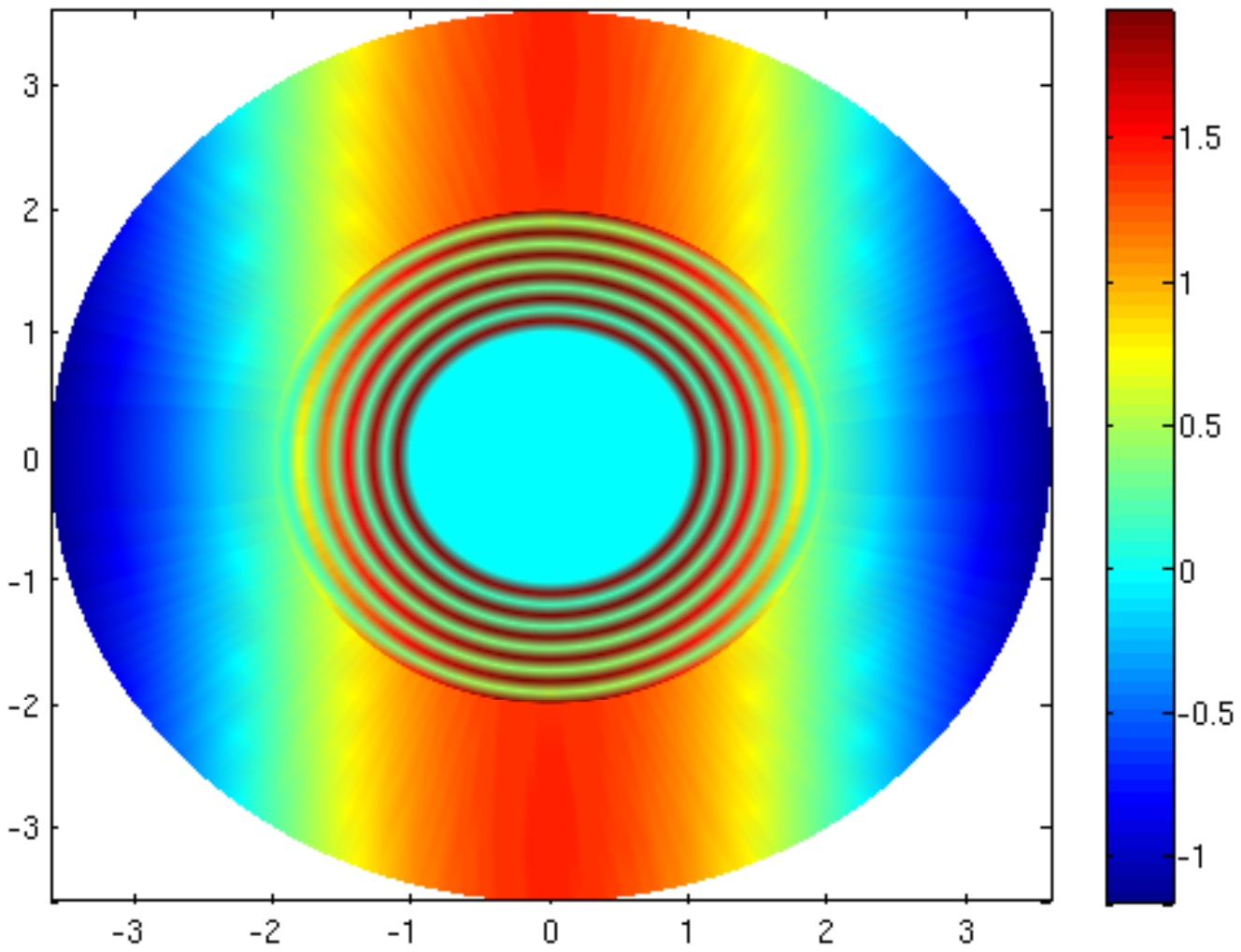}\ \ 
\hspace{-3.7cm}
\includegraphics[width=0.83\linewidth]{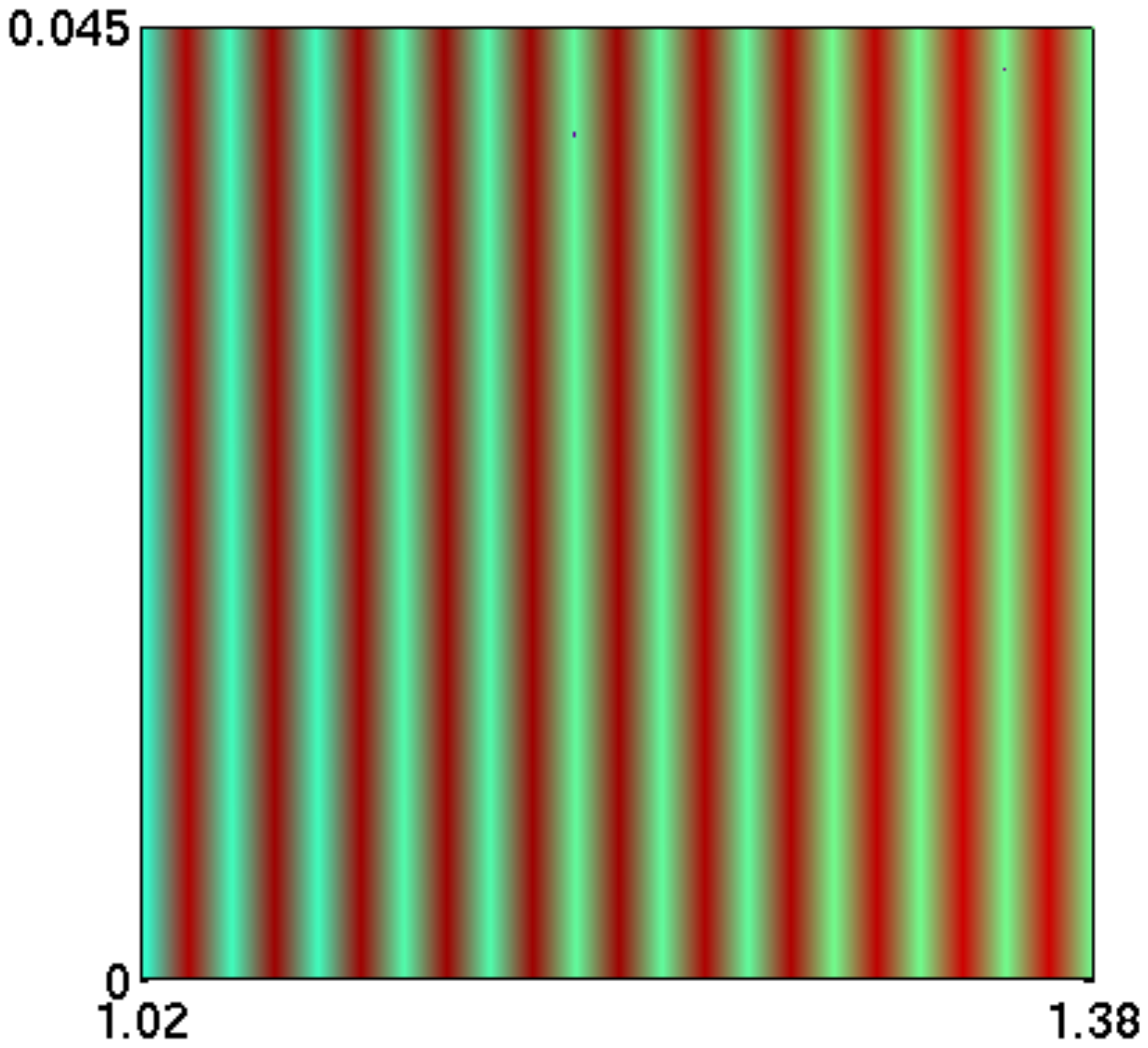}} 
\end{center}         
\vspace{-6.0cm}
\caption{ {\bf Scattering from cloak.} 
Left: The real part of  $\psi$ when a plane wave scatters
from an approximate cloak in the case when $E$ is not an interior eigenvalue.
Due to  limited resolution, $\psi$  is sparsely
sampled in radial direction; in reality, $\psi$ oscillates in the cloak more than is shown.   Right: A detail with finer resolution in the cloaking layers,  in polar coordinates.}
\end{figure}


\section{Three applications to quantum mechanics}\label{sec:sing qm}

In this section, we consider three examples of the results and ideas above to 
quantum mechanics.  Further discussion of applications is in \cite{GKLU7}.

\subsection{Case  study 1: Amplifying  magnetic potentials}

We first construct
a system consisting of a fixed homogeneous magnetic field and a sequence of
electrostatic potentials, the combination of which produce boundary or scattering
observations (at energy
$E$) making it appear as if the magnetic field blows up near a point.

The magnetic Schr\"odinger equation
with a magnetic potential $A$ (for  magnetic field  $B=\nabla\times A$) and electric potential $V$ is of the form
\beq \label{schr A}
-(\nabla+iA)^2 \psi +V \psi =E\psi,\quad \hbox{in }\Omega,
\quad \psi|_{\p \Omega}=f,
\eeq
where we have added the Dirichlet boundary condition on $\p \Omega$. 
Take now $V= V_n^E$ and denote the corresponding solutions of (\ref{schr A})
by $\psi_n$. Let $u_n:=\gamma_n^{-1/2}\psi_n$; then these $u_n$ satisfy, cf.
(\ref{gauge})--(\ref{24.4.1}), 
\ba
-g_{n}^{-1/2} \nabla_A \cdotp\gamma_{n} \nabla_A u_n=Eu_n,
\quad u_n|_{\p \Omega}=f,
\ea
where $\nabla_A:=\nabla+iA$. Similar to the considerations above, we see that if $n \to \infty$, then
$u_n \to u$, where $u$ is the solution to the problem
\ba
-g^{-1/2} \nabla_A \cdotp\sigma \nabla_A u=Eu,
\quad u|_{\p \Omega}=f.
\ea
Letting $w(y)=u(x)$ with $x=F(y),\, y \in B_3 \setminus\{{\it O}\},\,
x \in \Omega \setminus B_1$, we have that $w$ is the solution to the
magnetic Schr\"odinger equation, at energy $E$,  with $0$ electric potential
and magnetic potential ${\tilde A}$
\ba
-(\nabla+i{\tilde A})^2 w-Ew=0, \quad \hbox{in} \,\, B_3.
\ea
Since magnetic potentials transform as differential $1-$forms, we see that, briefly using subscripts for the coordinates,
\ba
\tilde A_j(y)=\sum_{k=1}^3 A_k(x) \frac {\p x_k}{\p y_j}
\ea
Now take the linear magnetic potential $A=(0, 0, a x_2)$,  corresponding to  homogeneous magnetic field
$B=(a, 0, 0)$. By the transformation rule (\ref{transformation}),
${\tilde A}=A$ in $ B_3 - B_2$, while in $B_2$
\ba
{\tilde A}(y) = a \left( 1 +\frac{|y|}{2}\right) \frac{y_2}{|y|^4}
\left(-y_1 y_3,\,- y_2 y_3, \, (y_1)^2+(y_2)^2 +|y|^3/2 \right).
\ea
>From this we see that ${\tilde A}(y)$ blows up near $y=0$ as ${\it O}({|y|^{-1}})$
so that the corresponding magnetic field  ${\tilde B}(y)$ blows up near $y=0$ as
${\it O}(|y|^{-2})$.

Consider now the Dirichlet-to-Neumann operator for the magnetic
Schr\"odinger equation (\ref{schr A}) with $V=V_n^E$,
i.e., the operator $\Lambda_{\overline{V_n},A}$ that maps
\ba
\Lambda_{\overline{V_n},A}: \psi|_{\p \Omega} \mapsto \p_\nu \psi|_{\p \Omega}.
\ea
Then the above considerations show that, as $n \to \infty$,
$\Lambda_{\overline{V_n},A}f \rightarrow\Lambda_{0,\tilde A}f$. In other words, 
as $n \to \infty$,  the boundary observations
at energy $E$, for the magnetic
Schr\"odinger equation with a linear magnetic
potential $A$, in the presence of the large
electric potentials ${V_n^E}$, appear as those  of
a very large magnetic potential $\tilde{A}$ blowing up at the origin, in the presence of very small electric potentials.

\subsection{Case study 2: Almost trapped states concentrated in the
cloaked region.}\label{sec almost trapped}

Let $Q\in C^\infty_0(B_1)$ be a real potential. 
The magnetic Schr\"odinger equation (\ref{schr A}) with potential $V=Q+V_n^E$ is, after a gauge transformation,    \linebreak cf. Sec. 3.1, closely related to the operator 
\ba
D_nu=-g_n(x)^{-1/2}\nabla_A \cdotp\gamma_n \nabla_A u+Qu, 
\ea
with domain $\{u\in L^2(\Omega):D_n u\in L^2(\Omega), u|_{\p \Omega}=0\}$.
We also define the operator 
$D$,
\ba
Du=-g(x)^{-1/2}\nabla_A \cdotp\sigma \nabla_Au+Qu,
\ea
which is a selfadjoint operator in the weighted 
space $L^2_g(\Omega)$
with an appropriate domain related to
the Dirichlet boundary condition $u|_{\p \Omega}=0$.
The operators $D_n$ converge  to $D$  (see \cite{GKLU6} for details) so that in particular 
for all functions $p$ supported in $B_1$
\beq
\label{resolvent}
\lim_{n\to \infty}(D_n-z)^{-1} p = (D-z)^{-1} p \quad \hbox{in}\,\,
L^2_g,
\eeq
if $z$ is not an eigenvalue of $D$.

\begin{figure}[htbp]
\vspace{-7.8cm}
\begin{center}
\includegraphics[width=1.2\linewidth]{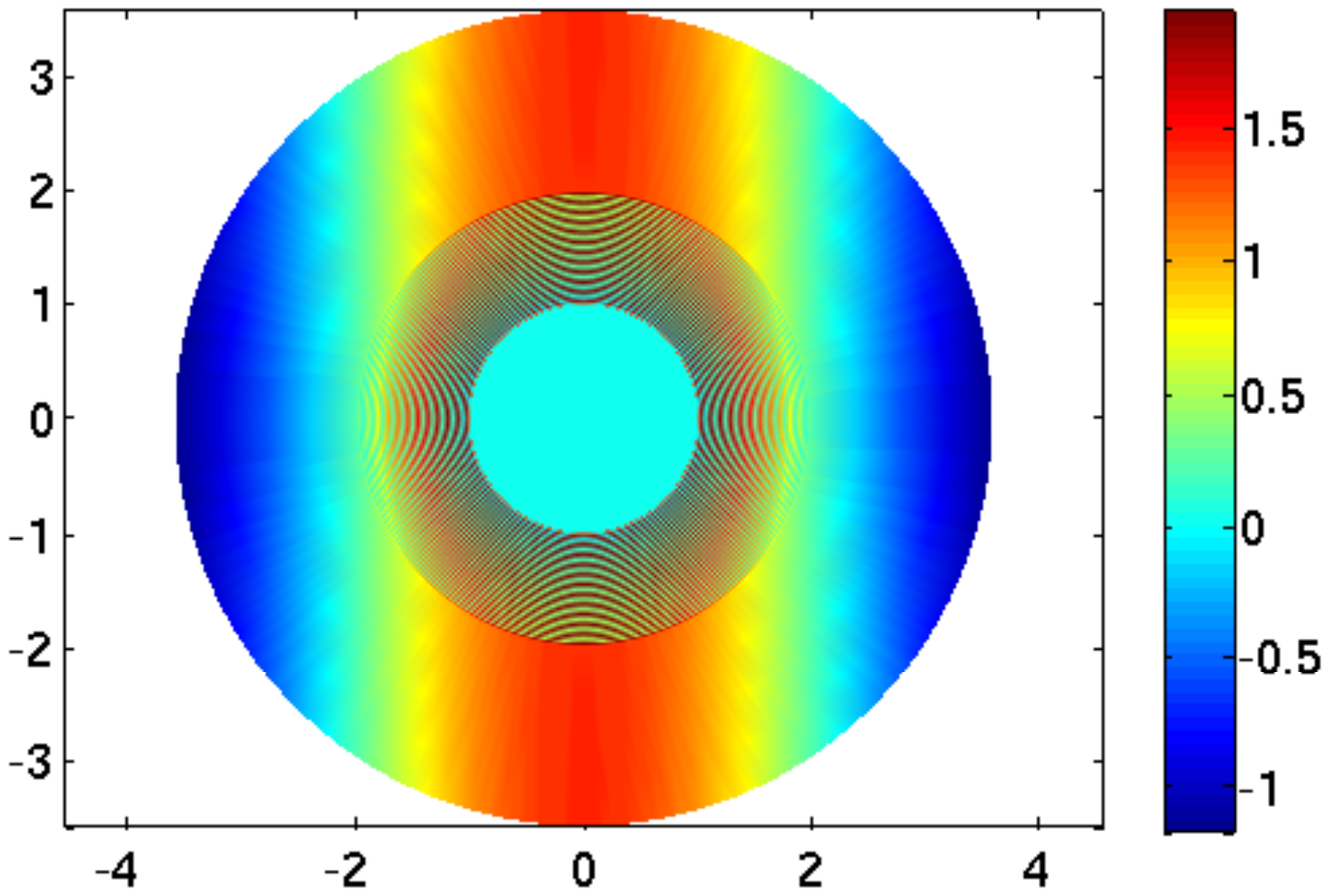}
\vspace{-7.0cm}
\end{center}
\caption{{\bf Plane wave and approximate cloak.} Re$\, \psi$ when $E$ is not an 
interior Neumann  eigenvalue: matter wave passes cloak 
almost unaltered. The Moir\'e pattern is an artifact.}
\end{figure}

\begin{figure}[htbp]
\vspace{-7.8cm}
\begin{center}
\includegraphics[width=1.2\linewidth]{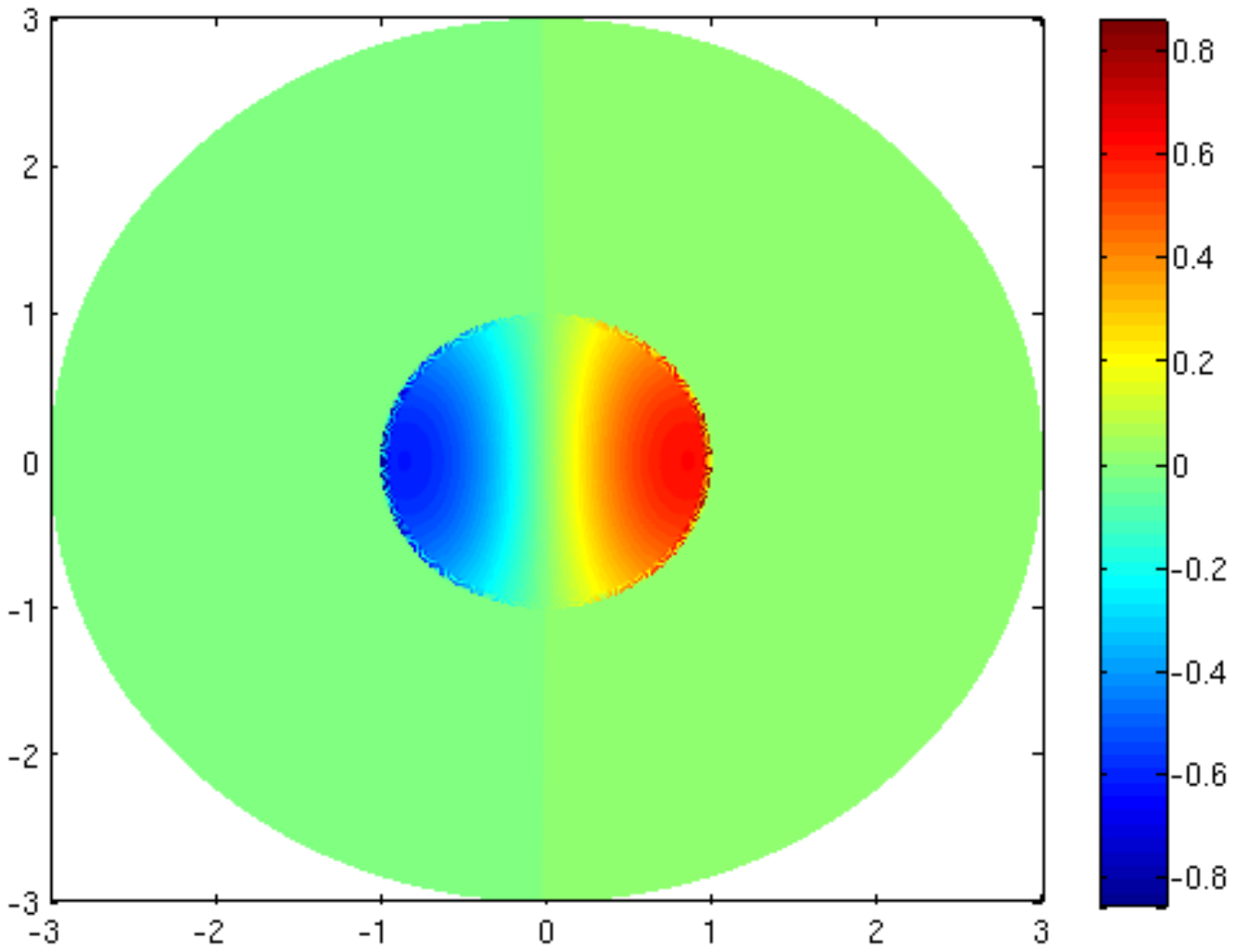}\qquad
\vspace{-7.0cm}
\end{center}
\caption{{\bf Almost trapped state.} A noncentral almost trapped
state: \qquad Re$\, \psi$ for potential
$W(x)=v_{in}\chi_{[0,0.9]}(r)$, $v_{in}=-3.92$ and $E=0.5$,
surrounded by the approximate cloak described in Sec 3.5.}
\end{figure}

Assume now that $E$ is a Neumann eigenvalue of
multiplicity one of the operator $-\nabla^2_A+Q$ in $B_1$ but is 
{\it not} a Dirichlet eigenvalue of operator $-\nabla^2_{{\tilde A}}$ in $\Omega=B_3$. 
Using  formulae (\ref{u-formula 1})--(\ref{u-formula 3}), one sees that then
$E$ is a eigenvalue of $D$ of multiplicity one and the corresponding eigenfunction $\phi$
is concentrated in $B_1$, that is, $\phi(x)=0$ for $x \in \Omega \setminus B_1$.
Assume, for simplicity, that $\kappa =1$, and let
$p$ be a function supported in $B_1$ that satisfies
\ba
a_{p}=\int_{B_1}\,dx\,p(x)\phi(x)=\int_{\Omega}\,dx\,g^{1/2}(x) p(x)\phi(x) \neq 0.
\ea

If $\Gamma$ is a contour in $\C$ around $E$ containing only one eigenvalue of $D$,
then
\beq\label{eq Seattle}
 \frac 1{2\pi i}\int_\Gamma \,dz\,
(D-z)^{-1}p=a_{p}\phi.
\eeq
However, by (\ref{resolvent}),
\ba
\frac 1{2\pi i}\int_\Gamma \,dz\,
(D-z)^{-1}p = \lim_{n \to \infty} \frac 1{2\pi i}\int_\Gamma \,dz\,
(D_n-z)^{-1}p.
\ea 
By standard results from spectral theory,  e.g., \cite{Kato}, this implies that
if $n$ is sufficiently large then there is only one  eigenvalue $E_n$ of $D_n$ inside $\Gamma$,
and $ E_n \to E$ as $n\to \infty$. Moreover,
\ba
a_{p}\phi=
\lim_{n\to \infty} a_{n,p}\phi_n,
\ea
where  $\phi_n$ is the eigenfunction of $D_n$ corresponding to the eigenvalue $E_n$ and  $a_{n, p}$ is  given as
\ba
a_{n, p}=\int_{\Omega}\,dx\,g_n^{1/2}(x) p(x)\phi_n(x) =\int_{B_1}\,dx\,p(x)\phi_n(x).
\ea
This shows, in particular,  that, when $n$ is sufficiently large,
the
eigenfunctions $\phi_n$ of $D_n$ are close to the eigenfunction $\phi$
of $D$ and therefore are almost $0$ in $\Omega - B_1$.

Applying the gauge transformation
(\ref{gauge}), we see that the magnetic Schr\"odinger operator
$-\nabla^2_A +(V_n^{E_n}+Q)$  has $E_n$ as an eigenvalue,
\ba
-\nabla^2_A \psi_n+(V_n^E+Q)\psi_n= E_n \psi_n,
\ea
where $\psi_n=g_n(x)^{-1/2}\phi_n$.
It follows from the above that this eigenfunction  $\psi_n$
 is close to zero outside $B_1$.  This means that the corresponding
 quantum  particle  is mostly concentrated in  $B_1$, 
  which we may think of  as an almost trapped state  located in $B_1$.

\subsection{Case study 3: $\S^3$ quantum mechanics in the lab }

The basic quantum cloaking construction outlined above can be modified to 
make the wave function on $B_1$ behave (up to a small error) as though it were 
confined to a compact, boundaryless
three-dimensional manifold which has been ``glued" into the cloaked region. Mathematically, this could be any manifold, $M$, but for physical realizability,
one needs to take $M$ to be the three-sphere, $\S^3$, topologically, but not necessarily with its standard metric, $g_{std}$. By 
appropriate choice of a Riemannian
metric $g$ on $\S^3$, the resulting approximately cloaking potentials can be custom designed to support  an essentially arbitrary
energy level structure. 

\begin{figure}[htbp]
\vspace{-2.8cm}
\begin{center}
\includegraphics[width=0.7\linewidth]{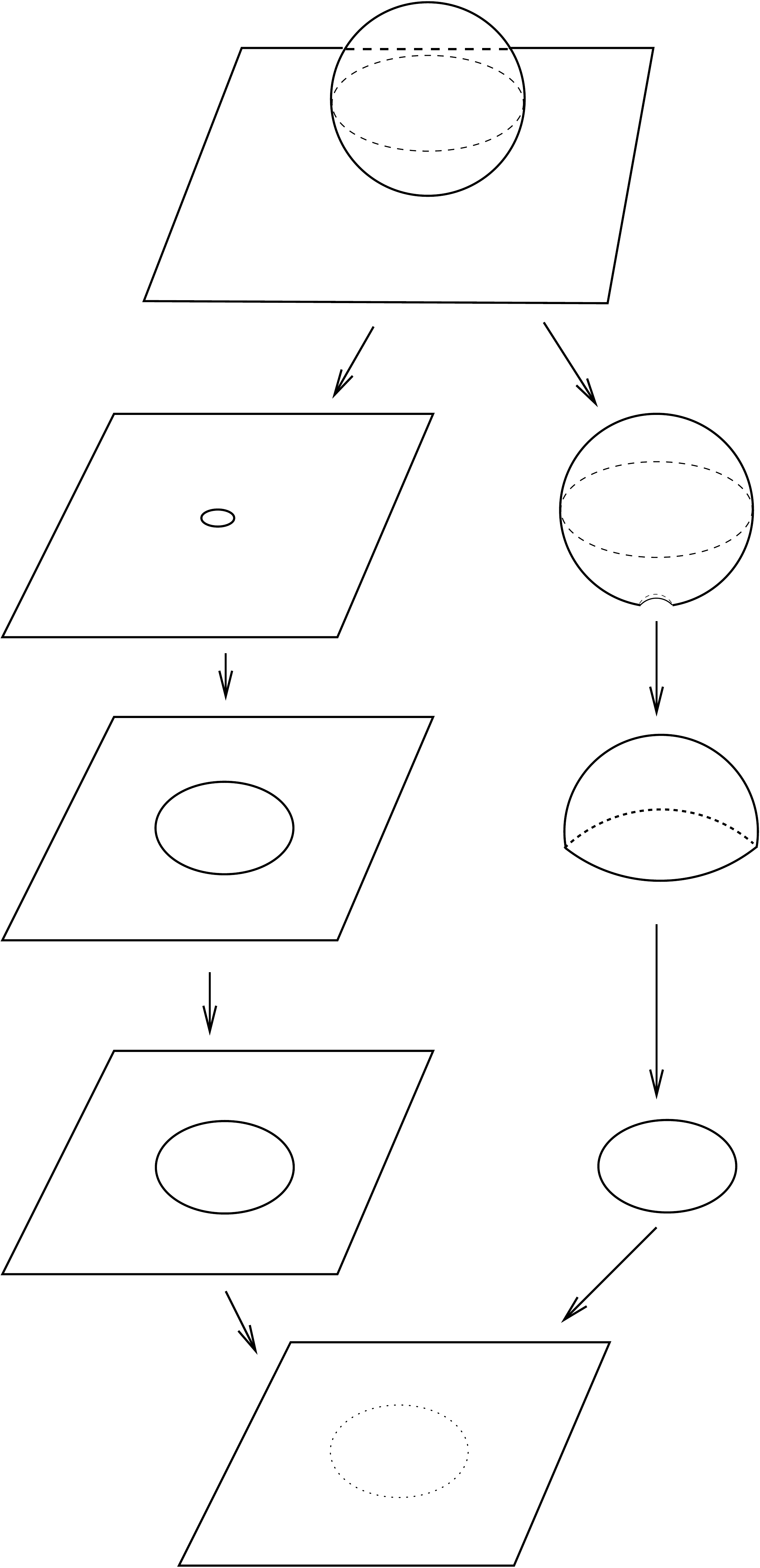}\qquad 
\end{center}
\caption{{\bf Schematic:}  Constructing an $\S^3$  approximate quantum cloak.}
\end{figure}

  As the starting point one uses not the original cloaking conductivity
$\sigma_1$  (the {\it single coating} construction), but instead  what was referred to in  \linebreak\cite[Sec. 2]{GKLU1}  as a {\it double} coating. 
This  is singular (and of course anisotropic) from {\it both} sides of $\Sigma$,
and in the  cloaking context corresponds to coating both sides of $\Sigma$ with appropriately matched metamaterials. 
Here,  we denote a double coating tensor  by $\sigma^{(2)}$. The part of such a $\sigma^{(2)}$ inside $B_1$  is specified by (i) choosing a Riemannian metric $g$ on
$\mathbb S^3$, with corresponding conductivity $\sigma^{ij}= |g|^\frac12 g^{ij}$; (ii) a small ball
$\tilde{B}_\delta$ about a distinguished point $x_0\in\mathbb S^3$;  (iii) a blow-up transformation
$T_1:\mathbb S^3-\{x_0\}\to \mathbb S^3-\tilde{B}_\delta$ similar to the $F$ used in the standard single coating
construction; and (iv)  a gluing transformation $T_2:\mathbb S^3-\tilde{B}_\delta\to B_1$, identifying the boundary
of
$\tilde{B}_\delta$ with the inner edge of the cloaking surface, $\Sigma^-$.   Then, 
$\sigma^{(2)}$ is 
defined as ${T_{2}}_* \left({T_{1}}_*\sigma\right)$ on $B_1$, and an appropriately matched single
coating on $B_3-B_1$ as before. 
This correpsonds to a singular Riemannian metric $g^{(2)}$ on $B_3$, with a two-sided conical singularity at $\Sigma$. One can show \cite[Sec. 3.3]{GKLU1} that the finite energy
distributional solutions of the Helmholtz equation $(\nabla_{g^{(2)}}+\omega^2) u=0$ on $B_3$ split into direct sums of
waves on $B_3-B_1$, as for $\sigma_1$, and waves on $B_1$ which are identifiable with
eigenfunctions of the Laplace-Beltrami operator  $-\nabla_g^2$  on  the compact, boundaryless Riemannian manifold $(\mathbb S^3,g)$, with eigenvalue $\omega^2$.

If
one takes $g$ to be the standard metric on $\mathbb S^3$, then the first 
excited energy level is
degenerate, with multiplicity 4, while a generic choice of  $g$ yields all  energy levels simple. On the other hand, it
is known that, by suitable choice of the metric $g$, any desired finite number of  energy levels and
multiplicities at the bottom of the spectrum can be specified  arbitrarily \cite{CdV}, allowing approximate quantum cloaks  to be built that model
abstract quantum systems, with  the energy $E$ having any desired  multiplicity.

\bibliographystyle{amsalpha}

\end{document}